\documentclass[11pt]{article}
\pdfoutput=1

\usepackage{euscript}
\usepackage{amssymb}
\usepackage{amsfonts}
\usepackage{amsbsy}
\usepackage{amsmath}
\usepackage{epsfig}
\usepackage{amsthm}
\usepackage{amscd}
\usepackage{amstext}

\usepackage[pdftex]{hyperref}

\def\ben{\begin{equation}}
\def\een{\end{equation}}
\def\half{{\textstyle{1\over2}}}
\def\qtr{{\textstyle{1\over4}}}
\let\a=\alpha  \let\g=\gamma  \let\e=\varepsilon
   \let\k=\kappa
     
 \let\t=\tau

\let\w=\omega \let\G=\Gamma

\let\pa=\partial
\def\be{\begin{equation}}
\def\ee{\end{equation}}
\def\ba{\begin{array}}
\def\ea{\end{array}}

\def\dalemb#1#2{{\vbox{\hrule height .#2pt
       \hbox{\vrule width.#2pt height#1pt \kern#1pt
               \vrule width.#2pt}
       \hrule height.#2pt}}}

\newcommand{\bea}{\begin{eqnarray}}
\newcommand{\eea}{\end{eqnarray}}
\newcommand{\tr}{{\rm tr} }

\def\R{{{\Bbb R}}}

\def\Z{{{\Bbb Z}}}

\def\N{{{\Bbb N}}}

\begin{document}

\begin{flushright}
NSF-KITP-09-144\\
arXiv:0908.1788 [hep-th]
\end{flushright}

\begin{center}

\vspace{1cm} { \LARGE {\bf Quantum oscillations and black hole ringing}}

\vspace{1cm}

Frederik Denef$^{\sharp,  \natural, \flat}$, Sean A. Hartnoll$^{\sharp, \natural}$ and Subir Sachdev$^{\sharp, \natural}$

\vspace{0.8cm}

{\it ${}^\sharp$ Department of Physics, Harvard University,
\\
Cambridge, MA 02138, USA \\}
\vspace{0.5cm}

{\it ${}^\natural$ Kavli Institute for Theoretical Physics, University of California, \\
Santa Barbara, CA 93106, USA \\}
\vspace{0.5cm}

{\it ${}^\flat $ Instituut voor Theoretische Fysica, U Leuven, \\
Celestijnenlaan 200D, B-3001 Leuven, Belgium \\}

\vspace{0.6cm}

{\tt  denef, hartnoll, sachdev @physics.harvard.edu} \\

\vspace{2cm}

\end{center}

\begin{abstract}

We show that strongly coupled field theories with holographic gravity duals at finite charge density
and low temperatures can undergo de Haas - van Alphen quantum oscillations as a function of
an external magnetic field. Exhibiting this effect requires computation of the one loop contribution
of charged bulk fermions to the free energy. The one loop calculation is performed using a formula
expressing determinants in black hole backgrounds as sums over quasinormal modes.
At zero temperature, the periodic nonanalyticities in the magnetic susceptibility as a function of the
inverse magnetic field depend on the low
energy scaling behavior of fermionic operators in the field theory, and
are found to be softer than in weakly coupled theories. We also obtain numerical
and WKB results for the quasinormal modes of charged bosons in dyonic black hole backgrounds,
finding evidence for nontrivial periodic behavior as a function of the magnetic field.

\end{abstract}

\pagebreak
\setcounter{page}{1}

\tableofcontents

\pagebreak

\section{Introduction}

\subsection{Beyond universality: one loop effects in holography}

One objective of applications of the holographic gauge/gravity correspondence \cite{Maldacena:1997re} to condensed matter physics is the characterisation of exotic states of matter. Recent works have begun to uncover a rich structure in strongly coupled theories with holographic gravity duals at finite charge density. Initial studies \cite{Hartnoll:2007ih, Hartnoll:2007ip} focused on hydrodynamic aspects at higher temperatures, while 
many interesting ground states have emerged in later studies at low temperatures.
When probed with charged scalar operators these theories can exhibit low temperature instabilities towards superconducting phases \cite{Gubser:2008px, Hartnoll:2008vx, Hartnoll:2008kx}. Equally interesting are the cases in which the finite density theory admits gapless charged scalar excitations but no superconducting instability \cite{Denef:2009tp}. When probed with charged fermionic operators, the response functions of the theory appear to indicate an underlying Fermi surface with non-Landau liquid excitations \cite{Lee:2008xf, Liu:2009dm, Cubrovic:2009ye, Faulkner:2009wj}.

The recent discoveries listed above lead to a seemingly paradoxical situation. The presence or absence of superconducting instabilities and Fermi surfaces is sensitive to the charge and mass of matter fields in the gravitational bulk spacetime \cite{Denef:2009tp, Faulkner:2009wj}. Equivalently, it is sensitive to the charge and scaling dimensions of low dimensional operators in the field theory. In contrast, the thermoelectric equilibrium and response properties of the theory are completely independent of these fields: they are universally determined by the Einstein-Maxwell sector of the bulk action. The fact that quantities such as the shear viscosity over the entropy density  \cite{Kovtun:2004de} or the electrical conductivity over the charge susceptibility \cite{Kovtun:2008kx} are identical for many distinct strongly interacting theories has been considered a robust prediction of sorts of applied holography. However, it seems unphysical that, for instance, the frequency dependent electrical conductivity should be independent of whether or not the theory has a Fermi surface. The latter property depends on the matter content of the bulk theory while the former does not.

We take the viewpoint that the emergence of a universal thermoelectric sector from theories with radically distinct bosonic or fermionic response is an artifact of the classical gravity (`large $N$' in field theory) limit. In the absence of symmetry breaking condensates or relevant perturbations of the theory, the minimal gravitational dual at finite density and temperature is the charged AdS-Reissner-Nordstrom black hole (see e.g. \cite{Hartnoll:2009sz}). In this background only the metric and Maxwell fields are nontrivial and all matter fields vanish. The fact that the Einstein-Maxwell sector does not source the matter fields is why the equilibrium and thermoelectric linear response of the theory can be universal and blind to the matter content. Beyond the classical limit, however, it is clear that this `decoupling' cannot continue to hold. All matter fields will run in loops and modify the gravitational propagator while all charged matter fields will modify the electronic properties of the theory. These effects are obviously small in the large $N$ limit, yet they may lead to qualitatively new physics. Furthermore, in any putative `real world' application of holographic techniques, the desired value of $N$ is unlikely to be large.

This paper will initiate the study of bulk one loop physics in applied holography. These are a more general set of `$1/N$' corrections that go beyond those captured by including higher derivative terms in the gravitational action (see e.g. \cite{Kats:2007mq, Brigante:2007nu, Anninos:2008sj, Buchel:2008vz}). In particular, we will be interested in effects that are not captured by local terms in an effective action, as they involve loops of light fields.

Consideration of quantum effects in the bulk potentially opens the Pandora's box of quantum gravity.\footnote{The `box' of Pandora's box is apparently a mistranslation of the Greek word `pithos' which refers to a large jar, often human-sized. As well as sickness and toil, the opening of the jar was also said to unleash Hope onto humankind.} Control of the ultraviolet properties of the quantum theory will ultimately require embedding computations into a consistent theory of quantum gravity. This may appear at odds with the `phenomenological' approach to applied holography (e.g. \cite{Hartnoll:2009sz}) in which one restricts attention to a minimal set of fields needed to capture the physics of interest. At the one loop level these problems do not arise. Indeed the one loop physics of quantum gravity was fruitfully explored well before the availability of ultraviolet finite string theories \cite{gibbonshawking}. The technical point that makes this possible is that functional determinants can be computed up to the renormalisation of a finite number of local couplings in the classical gravitational action. In particular, the nonlocal effects of interest to us are insensitive to the ultraviolet completion of the theory.

\subsection{Quantum oscillations as a probe of exotic states of matter}

In this paper we will consider the bulk one loop correction to the free energy due to charged matter.
Our primary objective is to study the free energy as a function of an external magnetic field. Magnetic fields are fundamental probes of matter at low temperatures. The quantum Hall effect and closely related de Haas-van Alphen quantum oscillations are examples of phenomena in which Landau level physics reveals important information about the finite density system, such as the presence of a Fermi surface. 

In recent years, experimental studies of quantum oscillations have had a profound impact on our understanding of a variety of correlated
electron systems. In the hole-doped cuprates, the observation  \cite{doiron,cooper,nigel,cyril,suchitra,louis,suchitra2} of quantum oscillations with a period indicative of `small' Fermi surfaces
has shown that the `large' Fermi surface Fermi liquid state at large doping must be strongly modified in the underdoped regime.
In the electron-doped cuprates, quantum oscillations with both small and large periods have been observed \cite{helm}, separated by a presumed quantum
phase transition. In these contexts, it appears of interest to catalog the states of matter which can exhibit the quantum oscillations,
apart from the familiar Fermi liquid suspects. Possible examples include fermionic matter coupled to gauge fields, or non-superfluid states of
bosons such as vortex liquids or `Bose metals'.

It is not yet clear to what extent we can interpret the finite density matter of the present gravity duals in terms of the concepts mentioned in the previous paragraph, but it is our hope that a study of quantum oscillations will advance our understanding of such issues. 
The classical bulk (large $N$) free energy is not manifestly written as a sum over Landau levels, as we shall see. The one loop correction to the free energy, in contrast, will naturally appear in this form. It follows that quantities such as the magnetic susceptibility can be expected to show novel qualitative features at the one loop level that are not visible classically.

In section \ref{sec:free} we review the computation of the low temperature magnetic susceptibility for free fermions and bosons with a finite chemical potential. The case of fermions leads to de Haas-van Alphen oscillations (at the low temperatures we consider, these can also be thought of as quantum Hall transitions). We then compute the leading order in large $N$ magnetic susceptibility in strongly coupled theories with gravitational duals in section \ref{sec:strong}, with no indication of quantum oscillations. We go on to consider the (bulk) one loop magnetic susceptibility in the strongly coupled theory. This is done using a, new to our knowledge, expression for determinants in black hole backgrounds written as a sum over the quasinormal modes of the black hole. The formula is derived in \cite{us} and allows us to use the recent analytic results of \cite{Faulkner:2009wj} on fermionic quasinormal modes.
Quantum oscillations are seen to appear from the one loop contribution of fermionic fields. We find that the periodic delta functions characterising the free fermion susceptibility at zero temperatures are replaced by power law divergences at strong coupling. In section \ref{sec:bosons} we numerically explore the quasinormal modes of charged bosonic fields, 
discussing the possibility of periodic oscillations due to bosons also.

\section{Free theories: fermions and bosons}
\label{sec:free}

To introduce some of the techniques and concepts we will use later, we first exhibit
the de Haas-van Alphen quantum oscillations in a more familiar setting. We will
consider the cases of
free charged bosons and free charged fermions in 2+1 dimensions. We work in Euclidean signature ($t = - i \tau$)
and place the theory in a background chemical potential $\mu$ and magnetic field $B$
\be
A = i \mu d\tau + B x dy \,.
\ee
We are interested in computing the free energy at low temperature ($T/\mu \to 0$)
as the magnetic field is varied at fixed chemical potential.

It is convenient to treat the case of bosons and fermions simultaneously. For this purpose we
can start with the Euclidean action for a complex scalar boson:
\be \label{sephi}
S_E[\Phi] = \int d^3x \Big[  |\partial \Phi - i q A \Phi|^2 + m^2 |\Phi|^2 \Big] \,,
\ee
and the following action for fermions
\be
S_E[\Psi] = \int d^3x \Big[  \overline \Psi \Gamma \cdot \left(\partial - i q A \right) \Psi + m \overline \Psi \Psi \Big] \,.
\ee
These two actions give the free energy
\be\label{eq:trlog}
\Omega_B = T \tr \log \left[- \hat \nabla^2  + m^2 \right] \,, \qquad \Omega_F = - \frac{T}{2} \sum_\pm \tr \log \left[- \hat \nabla^2  + m^2 \pm q B \right]
\ee
where $\hat \nabla_\mu =  \partial_\mu - i q A_\mu$.
The only important difference between the bosonic and fermionic cases is that the bosons are periodic in the thermal time circle whereas the fermions are antiperiodic. The extra term appearing in the fermionic case is the magnetic Zeeman splitting of the spin degeneracy. This term will not qualitatively affect the quantum oscillations.

The traces in (\ref{eq:trlog}) can be computed as a sum over eigenvalues of the Laplace operator. The eigenvalues are given by
\be
- \hat \nabla^2 \Phi + m^2 \Phi \pm q B \Phi = \lambda \Phi \,,
\ee
where the $\pm$ term should be added for fermions and is absent for bosons. We will retain this notation in the remainder of this section.
The eigenvalue spectrum can be determined exactly by separation of variables in this equation. Let
\be
\Phi = e^{- i \w_n \t + i k y} X_\ell(x) \,,
\ee
where $k \in \R$ and the thermal frequencies are
\bea
\w_n & = & 2 \pi n T \qquad \qquad \text{(bosons)} \,, \label{eq:bosons} \\
\w_n & = & 2 \pi (n+\half) T \quad \text{(fermions)} \,, \label{eq:fermionwn}
\eea
for $n \in \Z$. The $X_\ell(x)$ satisfy
\be\label{eq:aboveshift}
- X_\ell'' + q^2 B^2 \bar x^2 X_\ell = K_\ell X_\ell \,,
\ee
where we shifted the $x$ variable so that $\bar x = x - k/qB$.
This equation for $X_\ell$ is just the Schr\"odinger equation for a harmonic oscillator and therefore
\be
K_\ell = | q B |  ( 2 \ell + 1) \,,
\ee
with $\ell \in \Z^+\cup\{0\}$. The eigenfunctions are Hermite polynomials
$ X_\ell(\bar x) = e^{- | q B | \bar x^2/2} H_\ell(\sqrt{|qB|} \bar x) \,.$
Putting the above together leads to
\be\label{eq:lambda}
\lambda = m^2 + 2 | q B |  (\ell + \half \pm \half) - (i \w_n - q \mu)^2 \,.
\ee
We see that the eigenvalue $\lambda$ will be independent of the momentum $k$. This is the degeneracy of the Landau levels. We can now check that the degeneracy is in fact
\be\label{eq:degeneracy}
\int dk = \frac{|q B| A}{2 \pi} \,,
\ee
where $A$ is the 2 dimensional area of the sample. To see this suppose that we had a finite sample of size $L_x \times L_y$. The allowed values for the momentum would be $k = 2 \pi n_y/L_y$ for $n_y  \in \Z^+\cup\{0\}$. The shift $x \to x - k/qB$ we noted above is possible provided that $k/qB \leq L_x$. This places an upper bound on $n_y$ leading to (\ref{eq:degeneracy}).

Taking into account the degeneracy (\ref{eq:degeneracy}) of the Landau levels, one can perform the sum over eigenvalues to obtain the standard expressions for the free energy of bosons and fermions. For future comparison we will express the result in the following form
\be\label{eq:Om}
\fbox{$\displaystyle
\Omega_\text{free} = \pm \frac{| q B | A T}{2 \pi}  \sum_\ell \sum_{z_\star(\ell)} \log \left(1 \mp e^{-z_\star(\ell)/T} \right) \,.$
}
\ee
The upper sign is for bosons and the lower for fermions. For fermions one should additionally let $\sum_\ell \to \half \sum_{\ell^\pm}$, separating out the spin up and down contributions. A divergent temperature independent constant proportional to $\sum_\ell q \mu$ has been neglected. In the above result
\be\label{eq:zstar}
z_\star(\ell) = q \mu \pm \sqrt{m^2 + 2 | q B |  (\ell + \half \pm \half)}\,,
\ee
These values of $z_\star(\ell)$ are to be thought of as complex frequencies which give $\lambda=0$ upon analytic continuation $z = i \w_n$ of (\ref{eq:lambda}). That is to say, they are solutions to the equations of motion, and zeroes of
\be\label{eq:lambdaz}
\lambda(z,\ell) = m^2 + 2 | q B |  (\ell + \half \pm \half) - (z - q \mu)^2 \,.
\ee
Expressing the free energy as a sum over complex frequencies that give zero modes of the differential operator is the key step that we shall generalise below at strong coupling.

The sum over $\ell$ in (\ref{eq:Om}) will diverge. This is a temperature independent divergence, as can be seen by rewriting
\bea
T \sum_{z_\star(\ell)} \log \left(1 \mp e^{-z_\star(\ell)/T} \right) & = & T \left[  \log \left(1 \mp e^{-(\e_\ell-q\mu)/T} \right)+ \log \left(1 \mp e^{-(\e_\ell+q\mu)/T} \right) \right]  \nonumber   \\
 & & + \quad  \text{($T$ independent terms)} \label{eq:rewriting} \,.
\eea
The finite temperature sums over $\ell$ are now manifestly convergent.
We introduced the energy of the $\ell$th Landau level
\be
\e_\ell = \sqrt{m^2 + 2 | q B |  (\ell + \half \pm \half)} \,.
\ee
One can use a renormalisation method, such as zeta function regularisation, to control the zero temperature sums over Landau levels. At this point we should also comment on the zero magnetic field limit.
The $B \to 0$ limit is to be taken keeping
\be\label{eq:kB}
2 | q B | \ell \equiv k^2 \qquad \text{fixed as $B \to 0$} \,.
\ee
In this limit the sum over the Landau levels becomes an integral over momenta
\be
|q B| \sum_\ell^{\ell_\text{max.}} \to \int_0^{k_\text{max.}} k dk \,,
\ee
with $k_\text{max.}$ related to $\ell_\text{max.}$ via (\ref{eq:kB}). The difference between bosons and fermions due to Zeeman splitting drops out in this large Landau level limit.

From (\ref{eq:Om}) and (\ref{eq:rewriting}) we can see the de Haas-van Alphen magnetic oscillations in the case of fermions (the lower sign in these two equations). Take the $T \to 0$ limit of (\ref{eq:rewriting}) with fermionic signs and with $\mu$ and $B$ fixed. Whether or not a given term contributes to the sum over $\ell$ in this limit depends on whether $-\e_\ell \pm q \mu$ is positive or negative. If it is negative, then the exponential in (\ref{eq:rewriting}) diverges and the term gives a finite contribution. However, if it is positive, then the exponential goes to zero, the argument of the logarithm goes to one, and hence the total term goes to zero. Therefore we have
\be
\lim_{T \to 0}  \Omega = \frac{| q B | A}{2 \pi} \sum_\ell (q \mu - \e_\ell) \theta(q \mu - \e_\ell) + \cdots \,. \qquad (\text{fermions, }q\mu > 0) \label{eq:line2}
\ee
Here $\theta(x)$ is the Heaviside step function and is equal to 1 if $x > 0$ and zero otherwise. The dots denote analytic terms. As before for fermions $\sum_\ell \equiv \half \sum_{\ell^\pm}$.
We see that the free energy changes nonanalytically whenever one of the $z_\star(\ell)$ changes sign, say by tuning the magnetic field $B$. Note that this can only occur for one of the signs in (\ref{eq:zstar}), depending on the sign of $q \mu$. In (\ref{eq:line2}) we have assumed for concreteness that $q \mu > 0$. Of course, these nonanalyticities will get smoothed out at any finite temperature. The jumps in the derivative clearly occur whenever a Landau level crosses the fermi energy. To see the oscillations themselves we should differentiate twice to obtain the zero temperature magnetic susceptibility
\be\label{eq:freeoscillations}
\chi \equiv - \frac{\pa^2 \Omega}{\pa B^2} = - \frac{| q B | A}{2 \pi} \sum_\ell \frac{q^2 (\ell + \half \pm \half)^2}{\e_\ell^2} \delta(q \mu - \e_\ell) + \cdots \,,
\ee
where dots denote terms without delta functions. We can see that the susceptibility $\chi$ shows a strong response with period
\be
\Delta \left( \frac{1}{B} \right) = \frac{2 q}{q^2 \mu^2-m^2} = \frac{2 \pi q}{A_F} \,,
\ee
where $A_F = \pi k_F^2$ is the cross sectional area of the Fermi surface, with $k_F^2 = E_F^2 - m^2 = q^2 \mu^2-m^2$.

In (\ref{eq:line2}) the zero temperature free energy is piecewise linear in the chemical potential. It we compute the charge density via $\rho = \pa \Omega/\pa \mu$ then we find that the charge density is piecewise constant, with finite jumps at specific values of the magnetic field. These are the integer quantum Hall phases.

The boson system is quite different. The system is only stable if $\e_\ell > |q \mu|$. If $|q \mu|$ becomes larger than $\e_0$ then either the charged particles or antiparticles will condense, at any temperature. Using (\ref{eq:rewriting}) the expression (\ref{eq:Om}) is rewritten in the more familiar form
\be
\Omega = \frac{|q B| A T}{2 \pi} \sum_\ell \left[ \log \left(1 - e^{-(\e_\ell-q\mu)/T} \right)+ \log \left(1 - e^{-(\e_\ell+q\mu)/T} \right) \right] + \left. \Omega \right|_{T=0} \,. \qquad \text{(bosons)}
\ee
This last equation is recognised as the free energy of a gas of free charged particles and antiparticles. 
There are no jumps in the derivative, instead $\Omega$ diverges if $|q \mu|$ becomes equal to one of the $\e_\ell$.

\begin{figure}[h]
\begin{center}
\includegraphics[height=7cm]{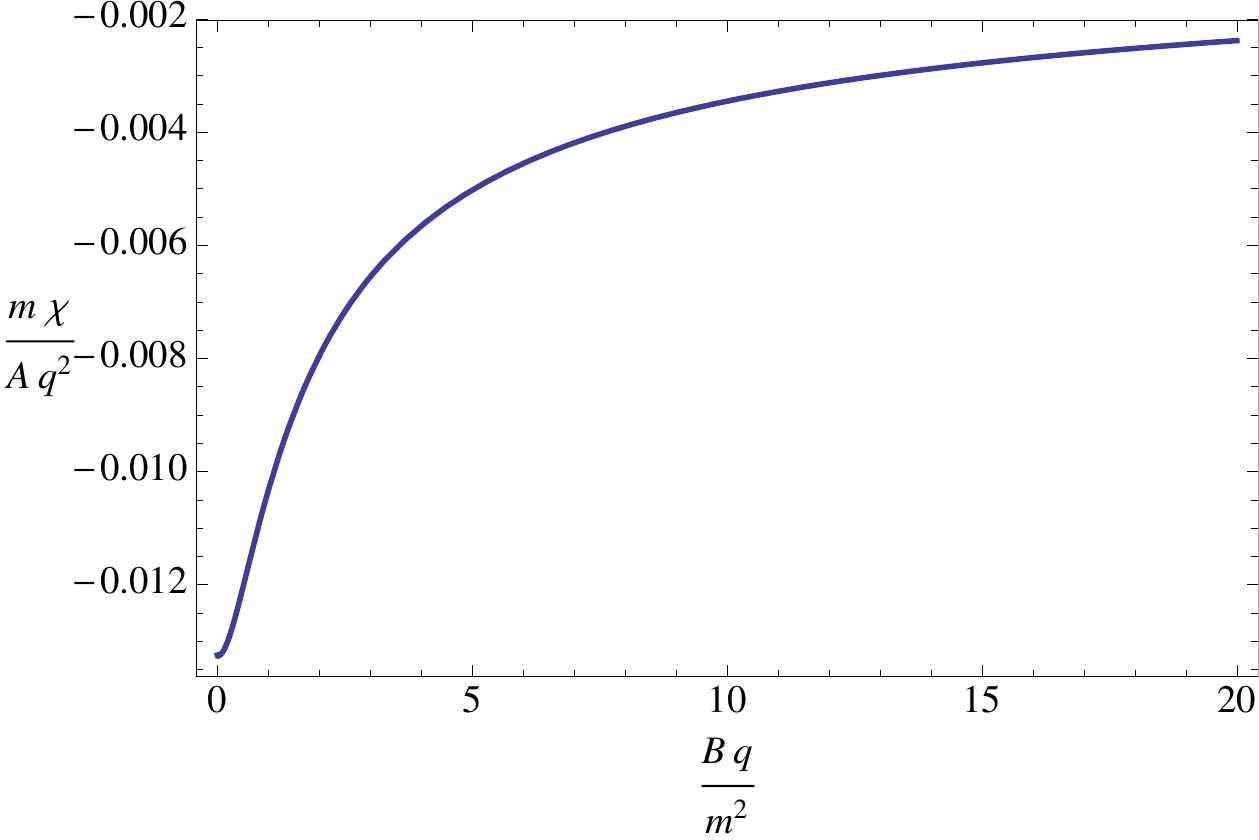}
\end{center}
\caption{The zero temperature magnetic susceptibility for bosons as a function of the magnetic field. The expression plotted has been made dimensionless by dividing by the sample area and multiplying by the boson mass $m$.}\label{fig:freebosons}
\end{figure}

Assuming that the mass is sufficiently large compared to the chemical potential so that the system is stable, the zero temperature free energy may be computed by, for instance, zeta function regularising the sum over Landau levels. One obtains
\be\label{eq:OmegaTzero}
\left. \Omega \right|_{T=0} = \frac{A |qB|^{3/2}}{\sqrt{2} \pi} \zeta_H\left(-\frac{1}{2},\frac{1}{2} + \frac{m^2}{2|qB|} \right) \,,
\ee
where the Hurwitz zeta function is defined by analytic continuation of
\be\label{eq:hurwitz}
\zeta_H(s,x) = \sum_{n=0}^{\infty} \frac{1}{(x+n)^s} \,.
\ee
The susceptibility obtained by differentiating this expression twice is shown in figure \ref{fig:freebosons}. There are clearly no oscillations of the type obtained for fermions. The values of the dimensionless susceptibility appearing in the plot are seen to be small. Note that the chemical potential does not appear in (\ref{eq:OmegaTzero}), so there is no charge density. The susceptibility in the plot is purely due to vacuum fluctuations.

In condensed matter applications, the theory Eq.~(\ref{sephi}) describes the superconductor-insulator transition of charged bosons
at integer filling in a periodic potential; for this case Eq.~(\ref{eq:OmegaTzero}) describes the diagmagnetic response
of the insulating phase.

\section{Strongly coupled theories with gravitational duals}
\label{sec:strong}

\subsection{The normal state geometry and large $N$ free energy}
\label{sec:normalstate}

In the previous section we reviewed the computation of magnetic susceptibility for free theories of bosons and fermions at finite chemical potential. We will now study the magnetic susceptibility of certain strongly coupled field theories, again with a finite chemical potential. Specifically, we study field theories which have large $N$ gravitational duals described `holographically' by Einstein-Maxwell theory in one dimension higher than the field theory (see e.g. \cite{Hartnoll:2009sz} for a motivation of this dual description). We work with 2+1 dimensional field theories and hence 3+1 dimensional gravitational duals.

Recall that our motivation is twofold. Firstly, we would like to see if any novel features arise in the magnetic response for theories that are stable against superconducting instabilities at finite chemical potential, despite having massless charged bosons \cite{Denef:2009tp}. Secondly, we would like to see if the putative Fermi surfaces identified in fermion spectral functions in \cite{Lee:2008xf, Liu:2009dm, Cubrovic:2009ye, Faulkner:2009wj} manifest themselves in the expected way as quantum oscillations.

In the absence of superconducting instabilities, the state of the field theory is dually described by a solution to Einstein-Maxwell theory. We are interested in thermodynamic properties and so we shall work in the Euclidean theory. The Euclidean action
is
\be\label{eq:einsteinmaxwell}
S_E[A,g] = \int d^{4}x \sqrt{g} \left[- \frac{1}{2 \kappa^2} \left(R + \frac{6}{L^2}
\right)  + \frac{1}{4 g^2} F^2 \right] \,.
\ee
Here $F=dA$ is the electromagnetic field strength.
The Einstein equations of motion are
\be\label{eq:EMeom}
R_{\mu\nu} - \frac{R}{2} g_{\mu\nu} - \frac{3}{L^2} g_{\mu\nu} = \frac{\kappa^2}{2 g^2}
\left(2 F_{\mu \sigma} F_{\nu}{}^{\sigma} - \frac{1}{2} g_{\mu\nu} F_{\sigma \rho} F^{\sigma \rho} \right) \,,
\ee
while the Maxwell equation is
\be
\nabla_\mu F^{\mu \nu} = 0 \,.
\ee

The normal state at a finite temperature, chemical potential and magnetic field is
described by the dyonic black hole metric (see e.g. \cite{Hartnoll:2007ai, Hartnoll:2009sz})
\be\label{eq:RNads}
ds^2 = \frac{L^2}{r^2} \left(f(r) d\tau^2 + \frac{dr^2}{f(r)} + dx^i dx^i \right) \,,
\ee
with
\be\label{eq:fmuB}
f(r) = 1 - \left(1 + \frac{(r_+^2 \mu^2 + r_+^4 B^2)}{\gamma^2} \right) \left(\frac{r}{r_+}\right)^3 +
\frac{(r_+^2 \mu^2 + r_+^4 B^2)}{\gamma^2} \left(\frac{r}{r_+}\right)^{4} \,,
\ee
together with the gauge potential
\be
A = i \mu \left[1 - \frac{r}{r_+} \right] d\tau + B x \, dy \,.
\ee
In these expressions we introduced the dimensionless quantity
\be
\gamma^2 = \frac{2 g^2 L^2}{\k^2} \,,
\ee
which is a measure of the relative strengths of the gravitational
and Maxwell forces. For a given theory, this ratio will be fixed. Some values arising in Freund-Rubin compactifications of M theory are described in \cite{Denef:2009tp}.

The field theory dual to this background has chemical potential $\mu$, magnetic
field $B$ and a temperature given by the Hawking temperature of the black hole
\be\label{eq:TBM}
T = \frac{1}{4 \pi r_+} \left(3 -  \frac{r_+^2 \mu^2}{\gamma^2} - \frac{r_+^4 B^2}{\gamma^2} \right) \,.
\ee
Note that whereas the chemical potential $\mu$ and temperature $T$ have mass dimension one in field theory, the background magnetic field has mass dimension two.
The free energy is given by evaluating the on shell classical action (see e.g. \cite{Hartnoll:2007ai, Hartnoll:2009sz})
\be\label{eq:Omega}
\Omega_0 = - \frac{A L^2}{2 \kappa^2 r_+^3} \left(1 + \frac{r_+^2 \mu^2}{\gamma^2} -  \frac{3 r_+^4 B^2}{\gamma^2} \right) \,,
\ee
where $A$ is the spatial area. From the free energy one computes the charge density
\be\label{eq:rho}
\rho = - \frac{1}{A} \frac{\pa \Omega_0}{\pa \mu} = \frac{2 L^2}{\k^2} \frac{\mu}{r_+ \g^2} \,,
\ee
and the magnetisation density
\be
m  = - \frac{1}{A} \frac{\pa \Omega_0}{\pa B} = - \frac{2 L^2}{\k^2} \frac{r_+ B}{\gamma^2} \,.
\ee
In these expressions, $r_+$ should be thought of as a function of $\mu,B$ and $T$ via (\ref{eq:TBM}). 

\begin{figure}[ht]
\begin{center}
\includegraphics[height=7cm]{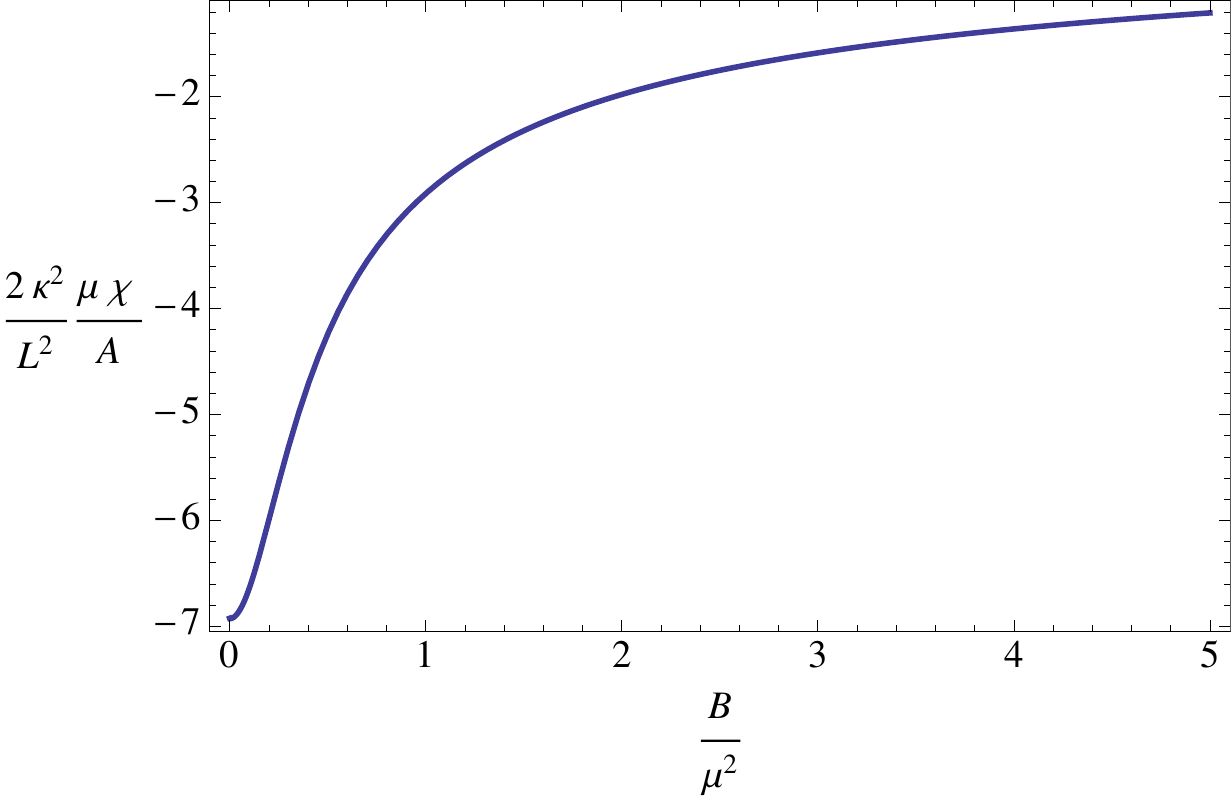}
\end{center}
\caption{The zero temperature magnetic susceptibility to leading order at large $N$ as a function of the magnetic field. The expression plotted has been made dimensionless by dividing by the sample area and multiplying by the chemical potential $\mu$.}\label{fig:chiclassical}
\end{figure}

Using the above results, the magnetic susceptibility $\chi = - \pa^2_B \Omega_0$ is easily computed from (\ref{eq:Omega}) and (\ref{eq:TBM}). The zero temperature result is plotted in figure \ref{fig:chiclassical}.

Figure \ref{fig:chiclassical} is the leading order large $N$ limit of the magnetic susceptibility.\footnote{We did not specify the connection between the normalisation of the action (\ref{eq:einsteinmaxwell}) and some dual field theoretical quantity $N$. In general one expects that $L^2/\kappa^2$ scales like $N$ to a positive power. That the coefficient of the classical action is large is precisely what allows the bulk side of the AdS/CFT correspondence to be treated classically.} The plot is disturbingly similar to that for free bosons in figure \ref{fig:freebosons}. Note however that the strongly coupled theory is scale invariant, and so the only scale at zero temperature is the chemical potential $\mu$, whereas in the free theory of the previous section we had a mass scale $m$.

\subsection{One loop ($1/N$) corrections to the free energy}

The leading order at large $N$ result for the free energy, (\ref{eq:Omega}), clearly does not show any nonanalytic structure as a function of the magnetic field at low temperature. The magnetic susceptibility is correspondingly uneventful as shown in figure \ref{fig:chiclassical}.

We will show in the remainder of this paper that this uneventfulness is an artifact of the large $N$ limit. As we mentioned in the introduction, a similar issue is known to arise in linear response. While the bulk Einstein-Maxwell theory captures all of the leading order in $N$ electromagnetic and thermal response of the field theory, it appears to be independent of the spectrum (charges and scaling dimensions) of low lying fermionic and bosonic operators in the theory. Yet it is precisely this spectrum that determines whether or not there is a superconducting instability  \cite{Denef:2009tp} and whether or not the fermionic response shows Fermi surface-like features \cite{Faulkner:2009wj}. A natural resolution to this tension is found in the fact that the Einstein-Maxwell and matter fields (fermions and bosons) are coupled in the bulk at a nonlinear level. Thus at higher orders in the $1/N$ expansion, or in higher point correlators, the matter fields will explicitly influence thermoelectric response.

In what follows we consider $1/N$ corrections to equilibrium thermodynamic quantities, in particular the magnetic susceptibility, which is simpler than considering linear response. We shall do this by computing one loop corrections to the classical result in the bulk. The flavour of the computation is identical to that for free fields in section \ref{sec:free}. The crucial difference is that the one loop contribution is to be computed in the curved black hole background of the previous subsection, which is 3+1 dimensional, as opposed to the 2+1 dimensions of the (strongly coupled) field theory.

There are several different sources of $1/N$ corrections to the free energy. It is helpful to identify those most likely to be related to the quantum oscillation structure we are seeking. The most universal one loop corrections to the free energy are those coming from the graviton and Maxwell field in (\ref{eq:einsteinmaxwell}). These will likely not lead to Landau-level related structure, however, as both fields are neutral. The same comment applies to higher derivative corrections to the classical action (\ref{eq:einsteinmaxwell}). Instead we will focus on the contribution of an additional
charged field, vanishing in the dyonic black hole background, which could be bosonic or fermionic.
For bosons the action takes the form
\be
S_E[\phi] = \int d^4x \sqrt{g} \Big[  |\nabla \phi - i q A \phi|^2 + m^2 |\phi|^2 \Big] \,,
\ee
while for fermions
\be
S_E[\psi] = \int d^4x \sqrt{g} \left[\bar \psi \Gamma \cdot \left(\pa + \qtr \w_{ab} \Gamma^{ab} - i q A \right) \psi + m \bar \psi \psi  \right] \,,
\ee
where $\w_{ab}$ is the spin connection. Roman letters denote tangent space indices.

There are one-loop contributions to the free energy
from fluctuations of the scalar and fermionic fields:
\be\label{eq:Z}
\Omega =  \, \Omega_0 + \Omega_B + \Omega_F = \, \Omega_0 + T \tr \log \left[- \hat \nabla^2  + m^2 \right]  - T \tr \log \left[\Gamma \cdot \hat D + m \right] + \cdots
\ee
where $\Omega_0$ is the classical result (\ref{eq:Omega}), $\hat \nabla = \nabla - i q A$ and $\hat D = \pa + \qtr \w_{ab} \Gamma^{ab} - i q A$. The boson and fermion masses in (\ref{eq:Z}) need not be the same, of course. The dots in (\ref{eq:Z}) indicate
that we are not computing the one loop contribution from the neutral fields $A$ and $g$. While the classical contribution $\Omega_0$ will scale as some positive power of $N$, the one loop logarithms in (\ref{eq:Z}) are order one. This is the sense in which we are computing a `$1/N$' effect.

\subsection{Determinants in black hole backgrounds and quasinormal modes}

In order to cleanly extract possible $T=0$ non-analyticities in the one loop determinants (\ref{eq:Z}), we would like to obtain an expression analogous to (\ref{eq:Om}) in the free field case. To do this, we must first write down the eigenvalue equation for bosons
\be
- \hat \nabla^2 \phi + m^2 \phi = \lambda \phi \,.
\ee
and for fermions
\be\label{eq:spinoreval}
\Gamma \cdot \hat D \psi + m \psi = \lambda \psi \,.
\ee
The next step is to separate variables. For bosons this is done by writing
\be\label{eq:phisep}
\phi = e^{- i \w_n \t + i k y} X_\ell(x) \phi (r) \,.
\ee
The quantities appearing in this expression are identical to those in section \ref{sec:free}. The important difference is that there is one more dimension, the bulk radial direction, and hence a new function $\phi(r)$.

Separation of variables is a little more complicated for spinors in a magnetic field because there are several components that couple differently to the field. However, it is straightforwardly achieved following Feynman and Gell-Mann \cite{Feynman:1958ty}. One introduces the auxiliary spinor $\chi$ defined by
\be
\psi = (\Gamma \cdot \hat D + \lambda - m) \chi \,,
\ee
which is found to satisfy the second order equation
\be\label{eq:secondorder}
- \hat D^2 \chi + \frac{1}{4} R \chi + \frac{i q}{2} F_{ab} \Gamma^{ab} \chi + (m - \lambda)^2 \chi = 0 \,.
\ee
This second order equation can now be separated exactly as in the bosonic case
\be\label{eq:chisep}
\chi = e^{- i \w_n \t + i k y} X_\ell(x) \chi (r) \,.
\ee
Every $\chi$ satisfying the second order equation (\ref{eq:secondorder}) gives an eigenspinor $\psi$ of the original Dirac operator (\ref{eq:spinoreval}). The solutions will be double counted, because of the extra derivatives. However, the matrix $\Gamma^5$ commutes with the second order operator. Therefore by imposing, say, $\Gamma^5 \chi = \chi$ one obtains the correct eigenfunctions without double counting.

As in section \ref{sec:free} above we will want to analytically continue $\w_n$ into the complex plane. Setting $z = i \w_n$ and substituting the separation of variables ansatze into the eigenvalue equations, we obtain `reduced' equations for each mode
\be\label{eq:radial2}
M_B(z,\ell) \phi = \lambda(z,\ell) \phi \,, \qquad M_F(z,\ell) \psi = \lambda(z,\ell) \psi \,.
\ee
These differential equations for $\phi(r)$ and $\psi(r)$, when viewed as eigenvalue problems, will provide a connection between $\lambda$ and $z$, similar to (\ref{eq:lambdaz}) in the free case. As previously, the $k$ momentum can be eliminated from the equations and only leads to the Landau level degeneracy. The important difference between (\ref{eq:radial2}) and (\ref{eq:lambdaz}) is that the $M_{B/F}$ are now differential operators in the radial direction, so we do not have an algebraic expression for $\lambda(z,\ell)$.

Mimicking the free theory procedure, the idea now is to express the determinants as sums over specific complex frequencies $z_\star(\ell)$ that lead to zero modes; $\lambda(z_\star(\ell),\ell) = 0$ solutions of (\ref{eq:radial2}). Because $M_{B/F}$ are differential operators we expect to find infinitely many such frequencies. For the operators to be well defined, we need to specify the boundary condition of $\phi(r)$ and $\psi(r)$ near the horizon at $r=r_+$ and near the boundary $r = 0$. The subtler boundary condition is at the horizon. The general radial behavior near the horizon is found to be
\be\label{eq:horizonbc}
\phi, \psi \sim (r - r_+)^{\alpha} + \cdots \,, \quad \text{with} \quad \alpha =  \pm \frac{i z}{4 \pi T} \,.
\ee
In computing the Euclidean determinant directly as a sum over eigenvalues, regularity at the Euclidean horizon requires taking
\be\label{eq:modw}
\alpha = \frac{|\w_n|}{4\pi T} \,.
\ee
This then shows that once we have defined the boundary condition for $M_{B/F}(z,\ell)$ on the imaginary $z=i \w_n$ axis, the positive and negative values of $\w_n$ will have different analytic continuations into the complex $z$ plane. It is important to treat this point carefully in deriving the formula we present shortly.

At general complex $z$ the two boundary conditions in (\ref{eq:horizonbc}) can be called ingoing (the minus sign) and outgoing (the positive sign). This corresponds to whether the corresponding Lorentzian signature solutions have flux going into the future horizon of the black hole, or coming out of the past horizon. On shell modes, with $\lambda(z_\star(\ell),\ell) = 0$, satisfying ingoing boundary conditions at the horizon are called quasinormal modes.

The quasinormal frequencies $z_\star$ of a wave equation in a black hole spacetime are poles in the corresponding retarded Green's function in the black hole background. To see this explicitly it is useful to consider the trace of the inverse of our operators $M_{B/F}$, which we will denote collectively as $M$. Starting on the imaginary axis we have
\be\label{eq:intG}
\tr \frac{1}{M(i \w_n,\ell)} = \int_0^{r_+} G(i\w_n,\ell,r,r) \, dr \,,
\ee
where the Euclidean Green's function satisfies
\be
M(i\w_n,\ell) \, G(i\w_n,\ell,r,r') = r^4 \delta(r,r') \,.
\ee
The expression (\ref{eq:intG}) follows directly from the usual representation of the Green's function as a sum over eigenfunctions. The boundary condition for the Green's function at the horizon is (\ref{eq:horizonbc}) together with (\ref{eq:modw}).

Now consider the analytic continuation of this trace to general complex $z = i \w_n$, where we analytically continue (\ref{eq:modw}) from the upper imaginary axis. That is, we take 
the minus sign (ingoing) boundary condition in (\ref{eq:horizonbc}). Denote this object by $\tr_- \frac{1}{M(z,\ell)}$. In general one needs to perform the integral in (\ref{eq:intG}) before analytically continuing. It is clear that the poles in this analytically continued Green's function with ingoing boundary conditions at the horizon are given by precisely the quasinormal frequencies of the black hole, as this is when $M(z,\ell)$ has a zero eigenvalue. See e.g. \cite{Ching:1995tj} for a more detailed discussion.\footnote{The quasinormal modes also give the poles of the retarded Green's function of the operator dual to the bulk field in the dual field theory  \cite{Son:2002sd}. The field theory Green's function is essentially given by the behaviour of our bulk Green's function near the boundary at $r=0$ \cite{Witten:1998qj}.}. As usual, continuing the Euclidean Green's function from the upper imaginary axis gives the retarded Green's function. If the black hole is stable against linearised perturbations (as they will be in the cases we study below) then these poles are necessarily in the lower half plane. Furthermore, at finite temperature, the quasinormal modes give isolated poles.

The conclusion of the previous paragraph is that the nonanalyticities of $\tr_- \frac{1}{M(z,\ell)}$
are isolated poles in the lower half $z$ plane. We could have instead analytically continued the Euclidean Green's function from the negative imaginary axis. Denote this object by $\tr_+ \frac{1}{M(z,\ell)}$. The $+$ boundary condition at the horizon corresponds to outgoing modes. This necessarily leads to the advanced Green's function, with poles in the upper half plane. In fact
\be\label{eq:conjugate}
\tr_+ \frac{1}{M(z,\ell)} = \overline{ \tr_- \frac{1}{M(\bar z,\ell)}} \,.
\ee
This relation follows from taking the complex conjugate of (\ref{eq:radial2}) and (\ref{eq:horizonbc}). In the following we will express our results in terms of the poles $z_\star$ of the retarded Green's function (the quasinormal modes), as these are more physical for most purposes. If we wish we can always obtain the poles of the advanced Green's function from (\ref{eq:conjugate}).

In the paper \cite{us} we derive the following formulae expressing the one loop contributions to the action coming from bosonic and fermionic determinants as a sum over the quasinormal modes of the operators $M_B$ and $M_F$ respectively. The reader may also find appendix \ref{sec:damped} useful, in which we derive an analogous formula for the simple case of a single damped harmonic oscillator. No assumption is made about the quasinormal modes forming a complete basis. For bosons
\be\label{eq:meisterformula}
\fbox{
$\displaystyle{
\Omega_B = - \frac{|q B| A T}{2 \pi}  \sum_\ell \sum_{z_\star(\ell)}  \log
\left( \frac{|z_\star(\ell)|}{4 \pi^2 T} \left| \G \left(\frac{i z_\star(\ell)}{2 \pi T} \right) \right|^2 \right) + \text{Loc} \,.}$}
\ee
For fermions we obtain
\be\label{eq:meisterfermions}
\fbox{
$\displaystyle{
\Omega_F = \frac{|q B| A T}{2 \pi}  \sum_\ell \sum_{z_\star(\ell)} \log \left( \frac{1}{2 \pi}
\left| \G \left(\frac{i z_\star(\ell)}{2 \pi T} + \frac{1}{2} \right) \right|^2 \right) + \text{Loc} \,.}$}
\ee
There difference between bosons and fermions is due to the different thermal frequencies (\ref{eq:bosons}) and (\ref{eq:fermionwn}).
In both of these two expressions, the Loc term refers to a `local' contribution to the one loop effective action for the metric and Maxwell fields induced by integrating out the charged bosons and fermions. We will discuss these terms a little more below, they will not contribute to the various interesting effects we are looking for. Finally, we should note that while we have written (\ref{eq:meisterformula}) and (\ref{eq:meisterfermions}) in a way adapted to Landau levels and magnetic fields, the representation of determinants in black hole backgrounds as sums over quasinormal modes is much more general \cite{us}. The formulae (\ref{eq:meisterformula}) and (\ref{eq:meisterfermions}) will be the strong coupling analogues of equation (\ref{eq:Om}).

Generally the sums over $\ell$ and $z_\star(\ell)$ in (\ref{eq:meisterformula}) and (\ref{eq:meisterfermions}) do not converge. These are high frequency divergences that should be renormalised, for instance using zeta function regularisation. Sometimes to control the asymptotic behavior it is useful to take a step back from the above expressions and reintroduce a sum over the thermal frequencies:
\be\label{eq:OmegaBH2}
\Omega_B = \frac{|q B| A T}{2 \pi}  \sum_\ell \sum_{z_\star(\ell)}
\left(\sum_{n \geq 0} \log \left| n + \frac{i z_\star(\ell)}{2 \pi T} \right|^2 - \log\left|\frac{z_\star(\ell)}{2 \pi T}\right| \right) + \text{Loc} \,.
\ee
An entirely analogous expression exists for fermions. This formula is related to the result (\ref{eq:meisterformula}) using the following identity from zeta function regularisation:
\be\label{eq:zetareg}
\sum_{n=0}^\infty \log(n+z) = - \left. \frac{d}{ds} \sum_{n=0}^\infty \frac{1}{(n+z)^s} \right|_{s=0}
= - \log \frac{\G(z)}{\sqrt{2 \pi}} \,,
\ee

Yet another expression for the determinant is in a spectral representation form. This is derived from (\ref{eq:OmegaBH2}) using contour integration and the fact that the $z_\star(\ell)$ are all in the lower half plane. For bosons we have
\be\label{eq:OmegaBH}
\Omega_B = \frac{|q B| A}{2 \pi}  \sum_\ell \sum_{z_\star(\ell)} \int^\infty_{-\infty} \frac{d\Omega}{\pi} \frac{1}{e^{\Omega/T} - 1} \text{Im} \, \log \left(z_\star(\ell)- \Omega \right) + \text{Loc}\,.
\ee
We will not develop this expression further. A rigorous treatment would need to address the validity of closing the contour and the divergences of (\ref{eq:OmegaBH}) at $\Omega=0$ and $\Omega=-\infty$. Regularity at $\Omega=0$ may impose constraints on the quasinormal modes. In appendix \ref{sec:damped} we show how this works for the case of a single damped harmonic oscillator. There is again an analogous integral expression for fermions.

Finally, we should say a few words about the `local' contribution $\text{Loc}$. Essentially $\text{Loc}$ contains the local UV counterterms as well as terms that ensure the correct large mass behavior in (\ref{eq:meisterformula}) and (\ref{eq:meisterfermions}). Equality of the right and left hand sides of these formulae at large mass, including order one terms, requires \cite{us} (for bosons, say)
\be\label{eq:local}
\text{Loc} = \left. \left\{T \tr \log \left[- \hat \nabla^2  + m^2 \right] +  
\sum_{z_\star}  \log
\left( \frac{|z_\star|}{4 \pi^2 T} \left| \G \left(\frac{i z_\star}{2 \pi T} \right) \right|^2 \right)\right\} \right|_{\Delta^{\geq 0}} \,.
\ee
Where $\left. \right|_{\Delta^{\geq 0}}$ means that we should only keep the terms which remain nonzero in the limit $\Delta \to \infty$ (the `nonpolar' terms). Here $\Delta$ determines the scaling of the field near the boundary, and is related to the mass by the standard AdS/CFT formulae
\be
\Delta (\Delta - 3) = L^2 m^2 \quad \text{(bosons)} \,, \qquad \qquad \Delta = \frac{3}{2} + L m \quad \text{(fermions, $m > -\half$)} \,.
\ee
To shorten the expression (\ref{eq:local}) we have written $\sum_{z_\star}$ to include the sum over the Landau levels and their degeneracy. The reason that $\Delta$ appears in (\ref{eq:local}) is that this is the quantity that determines the asymptotic boundary conditions. The proof in \cite{us} of the central formulae (\ref{eq:meisterformula}) and (\ref{eq:meisterfermions})  uses analyticity arguments in $\Delta$ rather than $m^2$.

The first term in (\ref{eq:local}) is closely related to the large mass limit of a determinant of the form Laplacian plus mass squared. It is well known, see e.g. \cite{Vassilevich:2003xt} for a review, that the only terms that survive the large mass expansion of such a determinant are given by integrals of local curvatures of the background metric and Maxwell fields. Therefore, the effect of this first term is to renormalise the Einstein-Maxwell action (\ref{eq:einsteinmaxwell}), including the generation of higher curvature terms. These terms are blind to Landau levels and therefore will not lead to nonanalytic physics as a function of the magnetic field.

The second term in (\ref{eq:local}) could likely be computed in principle by using WKB methods to obtain the quasinormal frequencies to the first few leading orders in a $1/\Delta$ expansion, perhaps along the lines of \cite{Festuccia:2008zx}. These WKB computations would not be expected to detect nonanalyticities of the sort we will describe shortly, which occur at low or zero frequencies. In the following we will therefore generally ignore the $\text{Loc}$ contribution to the determinant.

\subsection{Zero temperature nonanalyticities}
\label{sec:nonanalytic}

The zero temperature limit of (\ref{eq:meisterformula}) and (\ref{eq:meisterfermions}) is especially simple.
As our theory is scale invariant, only the ratios $B/\mu^2$ and $T/\mu$ are meaningful. Let us work at fixed $B/\mu^2$ and take the limit $T/\mu \to 0$. How do the quasinormal poles behave in this limit? The two possibilities for a given quasinormal frequency $z_\star$ are firstly that $z_\star \to 0$, for instance if $z_\star \sim T$, and secondly that $z_\star$ remains finite, which requires that $z_\star \sim \mu$. We will see explicitly in section \ref{sec:bosons} below that both possibilities occur. The quasinormal modes that go to zero with temperature coalesce and form a branch cut at zero temperature.

Formally taking the low temperature limit of (\ref{eq:meisterformula}) or (\ref{eq:meisterfermions}) gives
\be\label{eq:bhzeroT}
\lim_{T \to 0} \Omega_{B/F} = \pm \frac{|q B| A}{2 \pi}  \sum_\ell \sum_{z_\star(\ell)} \frac{1}{\pi} \text{Im} \, \left[ z_\star(\ell) \log \frac{i z_\star(\ell)}{2 \pi T} \right]  + \cdots\,.
\ee
In this expression the logarithmic branch cut must be taken along the positive imaginary $z$ axis. This is determined by the singularities of the gamma functions in (\ref{eq:meisterformula}) and (\ref{eq:meisterfermions}) which are along the positive imaginary $z$ axis. This zero temperature limit is discussed for the damped harmonic oscillator in appendix \ref{sec:damped}.

The sum in (\ref{eq:bhzeroT}) will only get finite contributions from modes that scale as $z_\star \sim \mu$ at low temperatures. 
Frequencies that go to zero with $T$ give a vanishing contribution, as is already discernable in (\ref{eq:meisterformula}) and (\ref{eq:meisterfermions}). However, the finite contribution can come from either isolated poles or those coalescing to give a branch cut: even though the coalescing poles eventually go to zero with $T$, at any finite $T$ there will be coalescing poles with $z_\star \sim \mu$.
For the poles forming a branch cut, the sum $\sum_{z_\star}$ in (\ref{eq:bhzeroT}) will become an integral.

In general the low temperature sum (\ref{eq:bhzeroT}) is still difficult to perform. One difficulty are the UV divergences in the sums. We will present in section \ref{sec:bosons} below some WKB results for the large frequency quasinormal modes that are a first step towards a direct evaluation of the UV tail of this formula. However, there are specific situations in which the representation as a sum of quasinormal modes becomes extremely useful. This is when a particular mode or set of modes undergoes nonanalytic motion as a function of a parameter such as $B/\mu^2$. Derivatives with respect to this parameter will then pick out the contribution of these particular modes as dominating over the others. Using results from \cite{Faulkner:2009wj} we will shortly perform the sum (\ref{eq:bhzeroT}) exactly over a set of poles close to the real frequency axis that undergo nonanalytic motion as a function of the magnetic field.

It was shown in \cite{Faulkner:2009wj} that quasinormal frequencies of charged fermions, i.e. $\lambda = 0$ solutions to the Dirac equation (\ref{eq:spinoreval}) with ingoing boundary conditions at the horizon, can undergo nonanalytic motion as a function of spatial momentum $k$. Specifically, if the charge of the fermion is big enough compared to its mass, $3 \, m^2 L^2 < q^2 \gamma^2$, then there exists a critical momentum $k=k_F$ at which a quasinormal mode bounces off the real frequency axis at $z=0$. This leads to a low energy peak in the spectral function of the dual field theory fermionic operator near to a particular finite momentum. At $T=0$ the peaks becomes a delta function. The momentum $k_F$ was therefore identified as the `Fermi momentum' indicative of an underling strongly coupled Fermi surface.

The results from \cite{Faulkner:2009wj}, at finite momentum but zero magnetic field, can be adapted to our context as follows.\footnote{Fermionic quasinormal modes in a magnetic field were recently studied in \cite{Albash:2009wz, Basu:2009qz}.} We can note that the magnetic field $B$ appears in the `second order Dirac equation' (\ref{eq:secondorder}) in two ways. Firstly it appears as just $B$ in the metric function $f(r)$ and in the spin-magnetic `Zeeman' interation $F_{ab} \Gamma^{ab}$. Secondly, it appears as $\ell B$, i.e. multiplied by the Landau level, in the gauge covariant kinetic term. If we take the limit $B \to 0$ with $2 \ell |q B| \equiv k^2$ fixed then we loose the first terms while retaining the kinetic term. As in the free field case discussed around (\ref{eq:kB}) above, this limit reproduces precisely the $B=0$ and finite momentum $k$ equation studied in \cite{Faulkner:2009wj}. We can therefore directly use results from that paper, with the pole now bouncing off the real axis at $2 \ell |q B| = k^2_F$. The $B \to 0$ with $\ell B$ fixed limit is not essential to use results from \cite{Faulkner:2009wj}. Keeping $B$ finite introduces some smooth $B$ dependence into the various `constants' that appear in this and the following sections.

\subsection{Summing low temperature poles: Quantum oscillations}

Before taking the strict zero temperature limit, it is useful to look at the pole motion at finite but low temperature.
At frequencies and temperatures that are small compared to the chemical potential, $z, T \ll \mu$, it is possible to solve the Dirac equation explicitly, see appendix D4 of \cite{Faulkner:2009wj}. Using the observation of the previous section we may translate the expressions from that paper into results for the quasinormal frequencies with a finite magnetic field in the limit $B \to 0$ with $\ell B$ fixed. As we noted, this limit is not essential but cleanly extracts the nonanalytic behavior.

A crucial parameter in the discussion of \cite{Faulkner:2009wj} is $\nu$. This quantity controls the low energy ($\w \ll \mu$) scaling dimension of the dual fermionic operator in the strongly coupled field theory. This scaling dimension
is related to the charge and mass of the field by
\be\label{eq:nu}
\nu = \frac{1}{\sqrt{12}} \sqrt{2 m^2 L^2 - q^2 \gamma^2 + \frac{3}{2} \frac{\gamma^2 k_F^2}{\mu^2}} \,.
\ee
The Fermi momentum in units of the chemical potential, $\gamma k_F/\mu$, also depends on $m$ and $q$. This dependence must be determined by numerically solving the Dirac equation in the Reissner-Nordstrom black hole background. A plot of $\nu$ as a function of $m$ and $q$ may be found in figure 6 of \cite{Faulkner:2009wj}. It can be shown that $\nu$ is always real.

We will assume for concreteness that $\nu < \half$ (a similar discussion will hold for the case $\nu > \half$). In this case, the quasinormal frequencies $z_\star$ in the low temperature and small frequency regime were found to be given by
\be\label{eq:smallw}
{\mathcal{F}}(z_\star) = 0 \,,
\ee
where
\be\label{eq:curlyF}
{\mathcal{F}}(z) = \frac{k_\perp}{\Gamma\left(\half + \nu - \frac{i z}{2 \pi T} - \frac{i q \gamma}{\sqrt{12}} \right)} -
\frac{h e^{i \theta} e^{i \pi \nu} (2 \pi T)^{2 \nu}}{\Gamma \left(\half - \nu - \frac{i z}{2 \pi T} - \frac{i q \gamma}{\sqrt{12}} \right)} \,.
\ee
We have rearranged the expression appearing in \cite{Faulkner:2009wj} because it will be important that ${\mathcal{F}}(z)$ has zeros but no poles. In (\ref{eq:curlyF}) we have introduced
\be\label{eq:propto}
k_\perp = \sqrt{2 \ell |q B|} - k_F \,,
\ee
which is a measure of the magnetic field and can be either positive or negative. The constants $h$ and $\theta$ in (\ref{eq:curlyF}) are determined in \cite{Faulkner:2009wj} in terms of the charge and mass of the fermionic field (numerically in the case of $h$). It will be sufficient for our purposes to take them to be order one in units of the chemical potential. The value of $\theta$ is constrained to lie in the range
\be\label{eq:range1}
0 < \theta < \pi (1 - 2 \nu) \,,
\ee
which guarantees that the poles are in the lower half plane for both signs of $k_\perp$. 

\begin{figure}[h]
\begin{center}
\includegraphics[height=6cm]{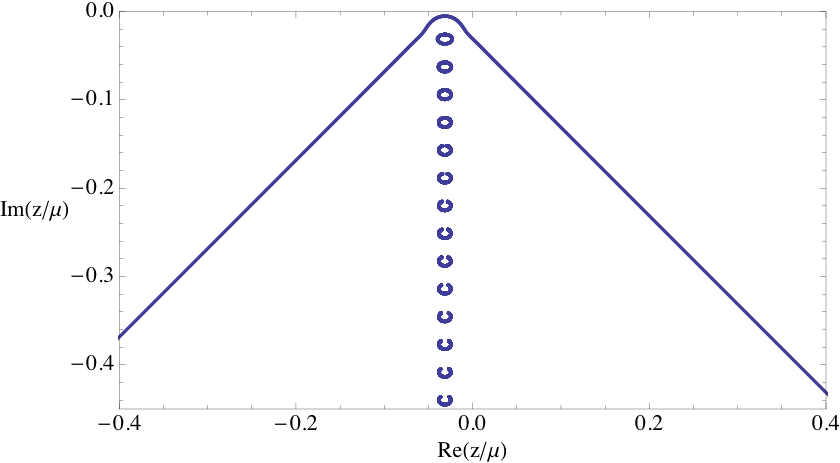}
\end{center}
\caption{Motion of the quasinormal frequencies closest to the real axis as $k_\perp/\mu$ is varied from $-1$ to $+1$, according to (\ref{eq:smallw}). The temperature is $T=0.005\mu$. The other constants are taken to have values $q=1, \gamma=\sqrt{12},\nu = 1/3,\theta=\pi/6,h=\mu^{1/3}$.}\label{fig:poledancing}
\end{figure}

The equation (\ref{eq:smallw}) will clearly lead to quasinormal frequencies of the form
\be
z^{(n)}_\star(k_\perp) = T {\mathcal{F}}^{(n)}\left(\frac{k_\perp \mu^{2\nu-1}}{T^{2\nu}} \right) \,,
\ee
for some sequence of functions ${\mathcal{F}}^{(n)}$. It is straightforward to solve (\ref{eq:smallw}) numerically and obtain the motion of the quasinormal poles as a function of $k_\perp$.\footnote{The authors of \cite{Faulkner:2009wj} have considered this problem in detail. We thank John McGreevy for drawing our attention to their appendix D4 and for sharing unpublished results on the motion of quasinormal poles in this low frequency regime.} In figure \ref{fig:poledancing} we show the low temperature motion of the poles closest to the real axis as $k_\perp$ is varied through zero, for a particular choice of numerical values of the parameters involved.

In figure \ref{fig:poledancing} we see several interesting effects. Firstly we can see the advertised pole that moves up to and then sharply bounces off the real axis. The bounce has been smoothed out at finite temperature.
Secondly, there are poles coalescing to form a zero temperature branch cut. These poles show a 
 nontrivial circular motion as a function of $k_\perp$. We now need to compute the magnitude of these effects on the magnetic susceptibility as $k_\perp$ goes through zero.

\begin{figure}[h]
\begin{center}
\includegraphics[height=5cm]{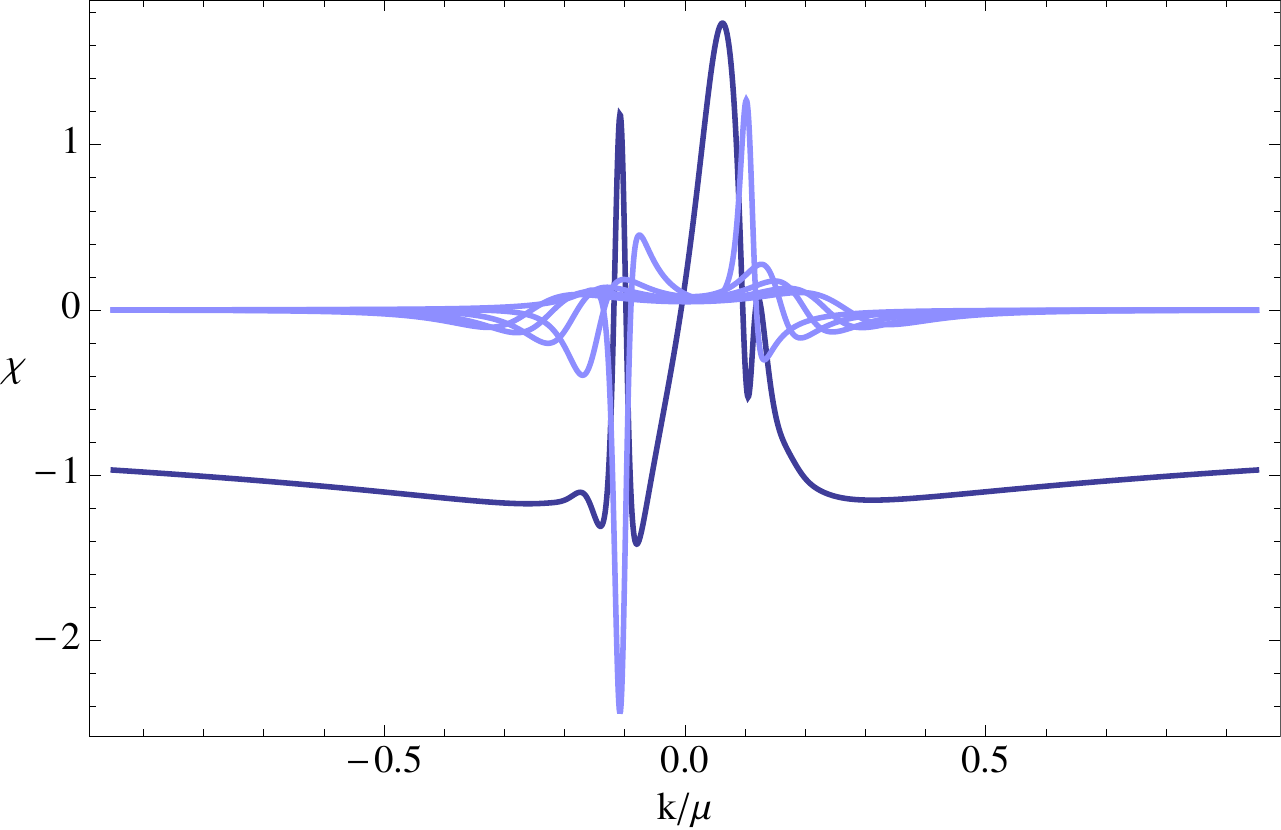}\includegraphics[height=5cm]{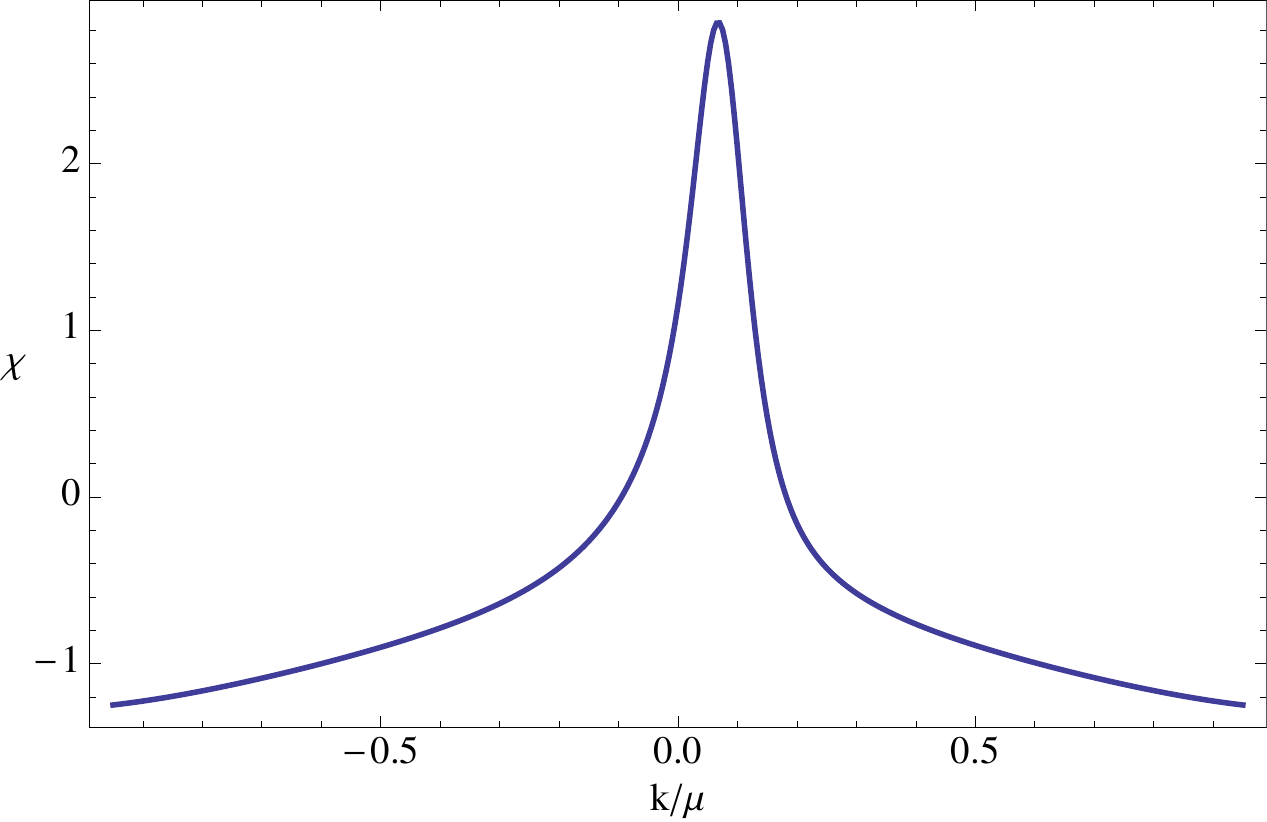}
\end{center}
\caption{Left: Contributions of the lowest few quasinormal modes to the magnetic susceptibility, according to (\ref{eq:meisterfermions}), as a function of $k_\perp/\mu$. The darker line is the pole nearest the real axis. Right: the total magnetic susceptibility due to the lowest fifty modes. The temperature is $T=0.005\mu$. The constants have the same values as in figure \ref{fig:poledancing}. The vertical axis is proportional to $\tilde \chi$ of (\ref{eq:chitilde}).}\label{fig:manychis}
\end{figure}

Figure \ref{fig:manychis} shows the contribution of these lowest few quasinormal poles to the magnetic susceptibility as a function of $k_\perp$. These are computed using our formula (\ref{eq:meisterfermions}) and strictly speaking we plot the quantity $\widetilde \chi$, see (\ref{eq:chitilde}) below, which is closely related to the susceptibility. The darker line in the first plot is the contribution of the `$T=0$' pole that bounces off the real axis in \ref{fig:poledancing}. The figure also shows the total susceptibility arising from the sum of the contributions of the lowest fifty modes. As anticipated, there is a strong feature in the response around $k_\perp=0$. We can also
discern other features in the individual responses of the modes. Somewhat magically, the motion of the `branch cut' poles is choreographed to precisely cancel out these extra features. The contribution of these other poles are the lighter lines in the first plot, while the second (right hand) plot shows the total response due to the lowest fifty poles. In the second plot only a single feature remains in the magnetic response. The cancellation between oscillations may perhaps be thought of as analogous to a Fourier transformation, in which sums of oscillations can cancel to give simple functions.

The peak seen in figure \ref{fig:manychis} will occur whenever $2\ell |qB| = k_F^2$. Thus the peaks are periodic in $1/B$ with period $2 \pi q/A_F$, as expected for quantum oscillations due to a Fermi surface. We will shortly make this statement sharper by going to the zero temperature limit.

The right hand plot in figure \ref{fig:manychis}, the sum of the lowest fifty poles, only makes sense if the series being summed is convergent. We are interested in the magnetic susceptibility, $\chi = - \pa_B^2 \Omega$, with $\Omega$ given by
(\ref{eq:bhzeroT}) and the quasinormal modes $z_\star$ given by (\ref{eq:smallw}). To determine convergence of this second derivative of the sum we need to know the dependence of $z_\star$ on the magnetic field $B$ at large values of $z_\star$.
Let us focus on $B$ close to the critical value $B_\ell$ at which the $\ell$th oscillation occurs: $B = B_\ell + \delta B = k_F^2/2\ell q + \delta B$. Then from (\ref{eq:propto})
\be\label{eq:kperp2}
k_\perp = \frac{\ell q}{k_F} \delta B + \cdots \,.
\ee
From (\ref{eq:smallw}), by expanding the right hand gamma function in (\ref{eq:curlyF}) in the vicinity of a negative integer, we find that at large $z_\star$ and for these small values of $k_\perp$:
\be\label{eq:dzdB}
\frac{d z_\star}{d B} \sim z_\star^{-2 \nu}\, .
\ee

We can now see by differentiating (\ref{eq:bhzeroT}) and using (\ref{eq:dzdB}) that while the sum over $z_\star(\ell)$ in
$\chi = - \pa_B^2 \Omega$ is UV divergent, this is only due to derivatives acting on the `trivial' overall factor of $B$ in (\ref{eq:bhzeroT}).
The divergent factor can be removed by considering for instance $- \pa_B^2 \Omega + 2 \pa_B (\Omega B^{-1})$, which does lead to a convergent sum. As we expect the term with most derivatives of $\Omega$ with respect to $B$ to capture the strongest nonanalyticities, extra terms depending on single derivatives of $\Omega$ should not be important for $k_\perp \sim 0$.
Alternatively we can define, using (\ref{eq:kperp2}),
\be\label{eq:chitilde}
\widetilde \chi \equiv - B \sum_\ell \frac{q^2 \ell^2}{k_F^2} \frac{\pa^2 \Omega_\ell}{\pa k_\perp^2} \,,
\ee
where $\Omega_\ell$ is the $\ell$th component of $\Omega = B \sum_\ell \Omega_\ell$. This leads to a convergent sum over $z_\star(\ell)$ and is again equivalent to
$\chi$ up to first derivatives of $\Omega$. In particular, we expect $\widetilde \chi \approx \chi$ at very low temperatures and $k_\perp \sim 0$. The finite quantity $\tilde \chi$ has been used as the vertical axis of figure \ref{fig:manychis}.

Having obtained convergent sums over $z_\star(\ell)$, the sum over $\ell$ itself is still not convergent. We shall not be concerned with this divergence, as we are considering low temperature nonanalyticities that occur for each $\ell$ individually at different values of the magnetic field. These nonanalyticities are not sensitive to the large $\ell$ UV divergences, analogously to the free field case of section \ref{sec:free}. In order to exhibit the quantum oscillations at high temperatures, one will likely have to perform the sum over $\ell$.

\subsection{The zero temperature limit}
\label{sec:zeroT}

In this section we will obtain the susceptibility exactly at zero temperature and for $k_\perp \sim 0$. To this end, the sum
over quasinormal modes in (\ref{eq:bhzeroT}) is helpfully rewritten as an integral along the real frequency axis
\be\label{eq:contoured}
\Omega_F = \frac{|qB| A}{2 \pi} \sum_\ell \frac{1}{\pi} \, \text{Im} \, \frac{1}{2 \pi i} \int_{-\infty}^{\infty} z \log\frac{i z}{2 \pi T} \frac{{\mathcal{F}}'(z)}{\mathcal{F}(z)} dz \,.
\ee
This expression follows from contour integration and the fact that ${\mathcal{F}}$, given in (\ref{eq:curlyF}), has zeros at the quasinormal modes $z_\star$ and no poles. The expression (\ref{eq:contoured}) is somewhat formal, but we now take two derivatives with respect to $B$ to obtain a convergent expression as described at the end of the previous section.

\begin{figure}[h]
\begin{center}
\includegraphics[height=6.5cm]{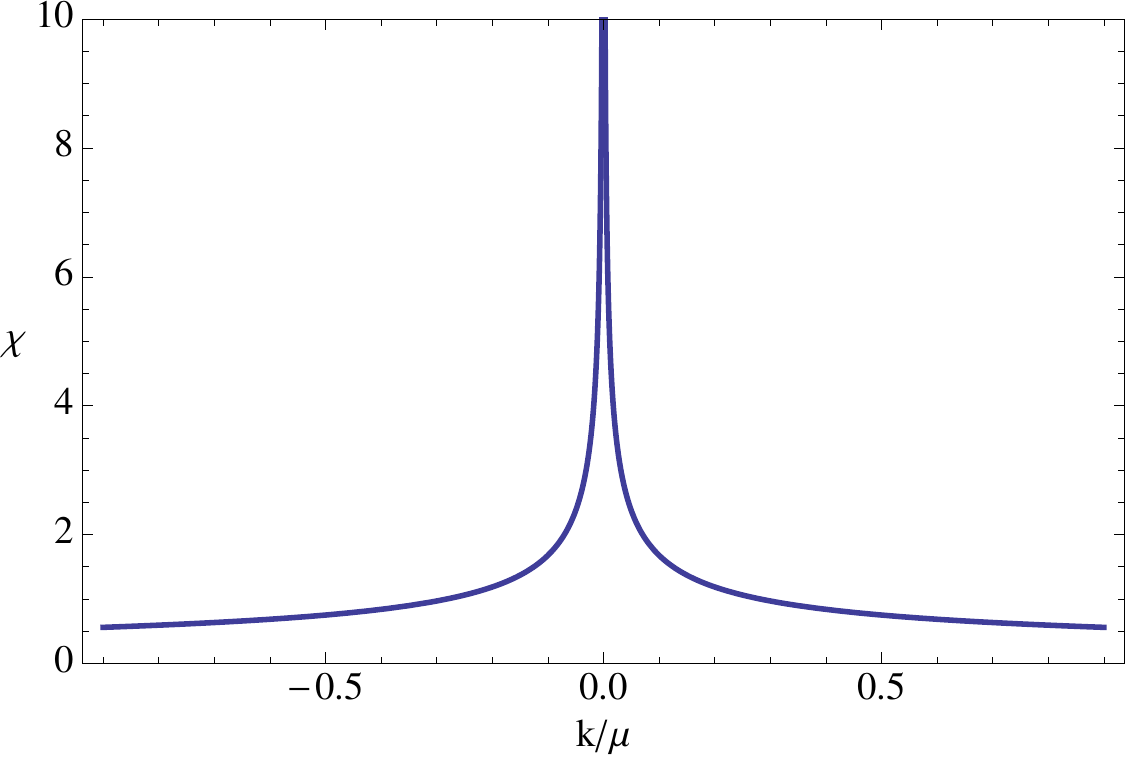}
\end{center}
\caption{The magnetic susceptibility at $T=0$ as a function of $k_\perp/\mu$.  The constants have the same values as in figure \ref{fig:poledancing}. The vertical axis is proportional to $\tilde \chi$ of (\ref{eq:chitilde}).}\label{fig:zerotchi}
\end{figure}

The zero temperature limit of the susceptibility is obtained by differentiating (\ref{eq:contoured}).
Using (\ref{eq:chitilde}) for the susceptibility (recall that $\widetilde \chi \approx \chi$ in this regime):
\be\label{eq:sus}
\left.  \widetilde \chi \right|_{T \to 0} = - \frac{|qB| A}{2 \pi} \sum_\ell \frac{q^2 \ell^2}{2 \pi^2 k_F^2} \text{Re}  \int_{-\infty}^{\infty} \frac{4 h \nu e^{i\theta} z^{2 \nu}}{\left(k_\perp - h e^{i\theta} z^{2 \nu} \right)^3}  \log \frac{i z}{2 \pi T} dz \,.
\ee
In the integrand it is important to take the branch cut due to the powers $z^{2 \nu}$ to run down the negative imaginary
axis, this is required by the coalescence of poles of the gamma functions in (\ref{eq:curlyF}). As we noted previously, the
logarithmic branch cut must run along the positive imaginary axis. The integral can be performed exactly to yield
\be\label{eq:dointegral}
\widetilde \chi  = \frac{|qB| A}{2 \pi}  \frac{(2 \nu - 1)}{4 \nu^2} \frac{q^2}{k_F^2 h^{1/2\nu}} \frac{\sin \frac{\theta}{2 \nu}}{\sin \frac{\pi}{2 \nu}}   \sum_\ell \frac{\ell^2}{(-k_\perp)^{2-1/2\nu}} \,, \qquad (k_\perp < 0) \,.
\ee
For $k_\perp > 0$ one replaces $k_\perp \to - k_\perp$ and $\sin \frac{\theta}{2 \nu} \to \sin \frac{\theta-\pi}{2 \nu}$. The integral is only convergent if $\qtr < \nu$. This extra condition is required to be able to close the contour in the lower half plane in the derivation of (\ref{eq:sus}) and is also the condition for the power of $k_\perp$ appearing in (\ref{eq:dointegral}) to be negative. For smaller values of $\nu$ one needs to differentiate the free energy more times to obtain a convergent integral and furthermore a divergent dependence on $k_\perp$. There is no temperature dependence in (\ref{eq:dointegral}). Technically this occurs because the integral in (\ref{eq:sus}) vanishes if the logarithmic term is not included. This shows that the $\log T$ in (\ref{eq:sus})  does not lead to a logarithmic divergence in the susceptibility at low temperatures.\footnote{Vanishing of the integral without the logarithm also indicates that $\sum_{z_\star} z_\star$, suitably regularised, is an analytic expression even though individual poles undergo nonanalytic motion. We suspect this may be a general phenomenon. Thus $\sum_{z_\star} z_\star \log z_\star$ is needed to extract the nonanalytic dependence on $B$. To obtain the correct answer one must sum all the poles near the real axis, it is not sufficient to focus on a single pole.} The zero temperature result (\ref{eq:dointegral}) is plotted in figure \ref{fig:zerotchi}.

Schematically, (\ref{eq:dointegral}) can be written as
\be\label{eq:chistrong}
\fbox{
$\displaystyle \chi = - \lim_{T \to 0} \frac{\pa^2\Omega_{F}}{\pa B^2} \sim +  |q B| A \sum_\ell \, \ell^2 \Big|2 \ell |q B| - k_F^2\Big|^{-2+1/2\nu} \,.$
}
\ee
The sign is important and physical. The divergences in the susceptibility at $2 \ell |q B| = k_F^2$ are seen to be positive, with opposite sign to the delta functions appearing for free fermions in (\ref{eq:freeoscillations}). The sign follows from the observation that $\sin \frac{\theta}{2 \nu}/\sin \frac{\pi}{2 \nu} < 0$ in the region we are studying: $\qtr < \nu < \half$ and (\ref{eq:range1}). The sum over $\ell$ can be performed in (\ref{eq:chistrong}) in terms of generalised zeta functions.
The formula (\ref{eq:chistrong}) is analogous to the result (\ref{eq:freeoscillations}) for free fermions. Once again, it indicates the existence of oscillations in the magnetic susceptibility with period
\be\label{eq:periodicity}
\Delta \left( \frac{1}{B} \right) = \frac{2 \pi q}{A_F} \,.
\ee

As well as the sign with which the susceptibility diverges, another importance difference with respect to the free fermion result (\ref{eq:freeoscillations}) is that the nonanalyticity is softer in the strongly coupled theory. Rather than the delta functions of the free theory (\ref{eq:freeoscillations}) we find the absolute value of a (generally non-integer) power in (\ref{eq:chistrong}). The power is determined by the low energy scaling dimension $\nu$ in (\ref{eq:nu}). Our computation is valid for $\qtr < \nu < \half$. These inequalities are satisfied for a range of values of $q$ and $m$, including for instance $m=0$ with $\gamma q=1$, see figure 6 of \cite{Faulkner:2009wj}.

To restate the main results of the last few sections
\begin{itemize}
\item For a range of values of the mass $m$ and charge $q$ of the bulk fermion there is a quasinormal pole which (at $T=0$) nonanalytically bounces off the origin of the real frequency axis at $2 \ell |q B| = k^2_F$.

\item For a certain range of $m$ and $q$, corresponding to $\qtr < \nu < \half$, these bounces produce periodic in $1/B$ divergences in the one loop magnetic susceptibility. The periodicity is given by (\ref{eq:periodicity}) and the strength of the divergence by (\ref{eq:chistrong}).
\end{itemize}
This behavior would seem to be aptly characterised as a strong coupling manifestation of de Haas-van Alphen oscillations at low temperatures. Thus our results simultaneously support the characterisation of $k_F$ as a Fermi momentum in  \cite{Lee:2008xf, Liu:2009dm, Cubrovic:2009ye, Faulkner:2009wj} and also indicate qualitative differences between de Haas-van Alphen oscillations at weak and strong coupling.

We now turn to a numerical study of the quasinormal modes of charged bosons and show that the modes can have an interesting magnetic field dependence in that case also. Quantum oscillations from stable charged bosons would be a novel effect.

\section{Charged scalar quasinormal modes of dyonic black holes}
\label{sec:bosons}

\subsection{Equations for bosons}

In the previous section we found analytic results for quantum oscillations due to fermionic quasinormal modes. This was possible because the relevant nonanalytic motion of the quasinormal mode occurred close to zero frequency, $z_\star \approx 0$, as the mode bounced off the real frequency axis. The Dirac equation in the zero temperature AdS-Reissner-Nordstrom black hole was solved analytically at small frequencies in \cite{Faulkner:2009wj}.

For bosons, in contrast, if a quasinormal mode moves towards $z_\star \approx 0$ it typically indicates the onset of a superconducting instability \cite{Gubser:2008px, Hartnoll:2008vx, Hartnoll:2008kx}. Rather than bounce back into the lower half frequency plane, the mode continues up into the upper half plane causing an instability and the condensation of the bosonic field. Therefore, if we wish to look for possible nonanalytic motion of bosonic quasinormal modes, without going through a phase transition, we will need to look away from small frequencies. To this end we will study the bosonic quasinormal modes numerically. The hunt for analogues of quantum oscillations leads us to look for special values of $K_\ell \sim \ell B$. Near the end of this section we will also look at the magnetic susceptibility at the onset of superconductivity.

Recall that after separating variables as in (\ref{eq:phisep}), and analytically continuing $z = i \w_n$, the eigenvalue equation became
\be\label{eq:MBeval}
M_B(z,\ell) \phi = \lambda \phi \,.
\ee
The `reduced' operator is found to take the form
\be\label{eq:M}
L^2 M_B(z,\ell) = - r^4 \frac{d}{dr} \left(\frac{f}{r^2} \frac{d}{dr} \right)  - \frac{r^2}{f} \left(z - q \mu \left(1- \frac{r}{r_+} \right)\right)^2 + \left(K_\ell r^2 +  (L m)^2 \right) \,.
\ee
Recall that $f$ was given in equation (\ref{eq:fmuB}).
In Appendix A we put this eigenvalue equation in Schr\"odinger form.
We are looking for the quasinormal modes of the operator $M_B$. These are $\lambda = 0$ eigenmodes of $M_B$ satisfying ingoing boundary conditions at the horizon and normalisable boundary conditions at infinity. We have already noted that ingoing boundary conditions at the horizon ($r=r_+$) corresponds to taking the minus sign in (\ref{eq:horizonbc}). Near the asymptotic boundary of the spacetime ($r=0$) the general behavior of $\lambda=0$ modes is
\be\label{eq:falloffs}
\psi = r^\alpha + \cdots  \,, \quad \text{with} \quad  \alpha = \frac{3}{2} \pm \sqrt{\frac{9}{4} +  L^2 m^2} \,.
\ee
Normalisability at infinity generally requires taking the faster of the two falloffs. In this paper we will
ignore the possible ambiguities that arise for masses sufficiently close to the Breitenlohner-Freedman bound ($m^2_{BF} = - 9/4L^2$) and simply impose the faster falloff at the boundary. For a discussion of determinants in AdS with $m^2_\text{BF} \leq m^2 \leq m^2_\text{BF}+1$ see e.g. \cite{Gubser:2002zh, Hartman:2006dy}.

\subsection{The matrix method for quasinormal modes}

The quasinormal modes of a black hole are complex frequencies
$z_\star(\ell)$ such that there are solutions $\phi$ satisfying
\be\label{eq:quasinormal}
M_B(z_\star(\ell),\ell) \phi = 0 \,,
\ee
together with ingoing boundary conditions at the horizon and normalisability at infinity. These are the frequencies that contribute to our sum (\ref{eq:meisterformula}).

The technical challenge we face is to find the quasinormal modes of a charged scalar field in planar dyonic Reissner-Nordstrom-AdS black holes. In particular, we will be interested in low and zero temperatures. While there is an immense literature on quasinormal modes, to our knowledge this particular problem has not been addressed. The most relevant references are collected in section 6.2 of the review \cite{Berti:2009kk}. It was noted in \cite{Horowitz:1999jd} that sometimes quasinormal modes in AdS are easier to find than in asymptotically flat spacetimes, because the AdS conformal boundary gives a regular singular point in the relevant differential equation rather than an irregular singular point. Unfortunately, the techniques of \cite{Horowitz:1999jd} will not work for us because at low temperatures there are singular points in the differential equation (\ref{eq:quasinormal}) that become arbitrary close to the horizon ($r=r_+$) and make a Taylor series expansion at the horizon useless. At strictly zero temperature the horizon becomes an essential singular point and a series expansion there has zero radius of convergence.

A useful discussion of asymptotically flat zero temperature Reissner-Nordstrom quasinormal modes can be found in \cite{Onozawa:1995vu}. The authors of that paper noted that, after a change of variables to bring infinity to a finite radial coordinate, then a Taylor series expansion at the midpoint between the horizon and infinity had a radius of convergence that reached both the horizon and infinity. The same property holds for our equation (\ref{eq:quasinormal}): the Taylor series about the midpoint $r_\text{mid} = \half r_+$ has radius of convergence $\half r_+$ and therefore reaches both the horizon at $r=r_+$ and the boundary $r=0$. This is true for all values of various parameters in the equation, including the zero temperature limit.

While \cite{Onozawa:1995vu} were then able to find the quasinormal modes by reducing a 5-term recurrence relation for the Taylor series about the midpoint to two 3-term recurrence relations and then using a continued fraction method due to Leaver \cite{leaver}, our case is more complicated. A Taylor series expansion of (\ref{eq:quasinormal}) about $r=\half r_+$ leads to a 9-term recurrence relation. Fortunately \cite{leaver} also presented a method for dealing with arbitrary length recurrence relations. We will now review the algorithm.

\begin{enumerate}

\item Expand $\phi$ in a series expansion about the midpoint, having first factored out the desired leading (singular) behavior at the horizon and infinity. Thus at finite temperature
\be
\phi = f^{- i z/4 \pi T} r^{\half (3 + \sqrt{9 + 4 L^2 m^2 })} \sum_{n=0}^{N} a_n (r - \half r_+)^n \,,
\ee
while at zero temperature
\be
\phi = e^{i z r_+^2/6 (r_+ - r)} f^{-i (4 z - 3 q \mu) r_+/36 } r^{\half (3 + \sqrt{9 + 4 L^2 m^2 })} \sum_{n=0}^{N} a_n (r - \half r_+)^n \,.
\ee
Note that at zero temperature $f(r) = 1 - 4 \left(r/r_+ \right)^3 + 3 \left( r/r_+ \right)^4$.

\item Plug the relevant series expansion into the differential equation (\ref{eq:quasinormal}) and expand. Collecting in powers of $r - \half r_+$ gives $N+1$ linear relations between the $N+1$ coefficients $\{a_n\}$. Write these as a matrix equation:
\be\label{eq:Amatrix}
\sum_{n=0}^N A_{m n}(z) \, a_n = 0 \,.
\ee
Because the recurrence relation between the $\{a_n\}$ involves nine terms in general, $A$ will have nonzero entries only along a diagonal band of width nine.

\item The quasinormal modes $z_\star$ are given by the zeros of the determinant of the matrix $A$
\be
\det A(z_\star) = 0 \,.
\ee
Given that the matrix $A$ is fairly sparse, this determinant can be numerically computed quickly and robustly using, for instance, {\sc Mathematica}.

\end{enumerate}

As an illustration and to introduce concepts we first present the results of this method for a neutral scalar field ($q=0$) with no magnetic field background ($B=0$) at low and zero temperature. For the moment we will make the choice of mass $m^2=0$; with this mass, neutral scalar fields are stable all the way down to zero temperature \cite{Denef:2009tp}. Furthermore, for concreteness we will take $\gamma=1$ throughout the remainder of this paper. For some values of $\gamma$ obtained via Freund-Rubin compactifications of M theory, see \cite{Denef:2009tp}.

\begin{figure}[h]
\begin{center}
\includegraphics[height=6.5cm]{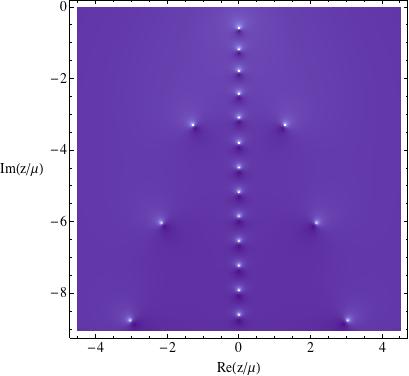}\hspace{0.2cm} \includegraphics[height=6.5cm]{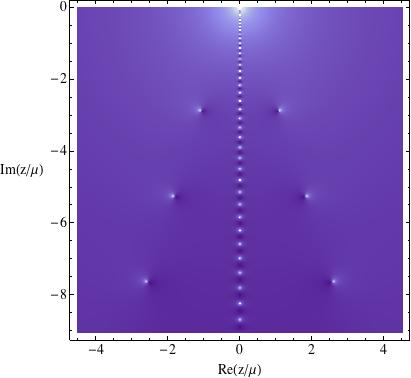}
\end{center}
\caption{Left: quasinormal modes at temperature $T/\mu = 0.075$. Right: quasinormal modes at zero temperature $T=0$. The plots show the lower half $z/\mu$ frequency plane, and bright spots denote quasinormal modes. Both plots have $q=0$, $m^2=0$, $\gamma=1$ and no magnetic field. The plots were generated as density plots of $\left| \det A'(z)/\det A(z) \right|$. The finite temperature plot has $N=200$ while the zero temperature plot has $N=500$. The discreteness of the poles on the imaginary axis at $T=0$ is an artifact of finite $N$. As $N \to \infty$ the poles coalesce and form a branch cut.}\label{fig:q0}
\end{figure}

Figure \ref{fig:q0} shows the quasinormal modes closest to the real axis for a small (left) and zero (right) temperature. The structure we are about to describe was to a large extent previously noted in \cite{Berti:2003ud, Wang:2004bv}. One sees clearly that at finite low temperature there are two distinct types of quasinormal modes. Along the negative imaginary axis we have a sequence of closely spaced modes, while on each side of the negative imaginary axis there is another sequence of modes descending diagonally. Because we are at low temperatures, it is natural to think of the closely spaced modes as having positions dominantly determined by $T$ whereas the off-axis modes are more sensitive to $\mu$. This statement can be made precise by varying the temperature, but we shall not go into detail here.
The symmetry of the plot under $z \mapsto - \bar z$ follows from the corresponding transformation of the differential equation (\ref{eq:quasinormal}) and ingoing boundary conditions (\ref{eq:horizonbc}) when $q=0$.

In the zero temperature limit, $T/\mu \to 0$, we should expect the modes along the negative real axis to bunch together and possibly form a branch cut. The right hand plot in figure \ref{fig:q0}, showing the zero temperature quasinormal modes, supports this picture. The fact that discrete poles are still visible in the plot is an artifact of truncating the differential equation to a finite matrix equation (with rank $N+1$) in step 2 of the algorithm we presented above. As $N$ is increased one can check that the modes in the figure move up the imaginary axis, becoming arbitrarily clumped as $N \to \infty$. The modes away from the imaginary axis remain fixed, consistent with the notion that their spacing is dominantly set by $\mu$.

For the case of a neutral scalar it is straightforward to argue analytically that there is a branch cut in the retarded Green's function at zero temperature (while the quasinormal modes correspond to poles in the Green's function). Making the change of variables
\be\label{eq:changevar}
\phi = r \Psi \,, \qquad \frac{dr}{f} = ds \,,
\ee
the differential equation (\ref{eq:quasinormal}) with $m^2=q=0$ at $T=0$ becomes the Schr\"odinger equation
\be\label{eq:q0schro}
- \frac{d^2 \Psi}{ds^2} + \frac{f(r)}{r} \left(\frac{2 f(r)}{r} - f'(r) \right) \Psi = \w^2 \Psi \,,
\ee
where we think of $r$ as $r(s)$. By integrating (\ref{eq:changevar}) one easily finds that $s \to + \infty$ as $r \to r_+$. Furthermore, the potential in the Schr\"odinger equation (\ref{eq:q0schro}) has the leading near-horizon behavior
\be
V = \frac{f(r)}{r} \left(\frac{2 f(r)}{r} - f'(r) \right) = \frac{r_+}{3 s^3} + \cdots \quad \text{as} \quad s \to \infty \,.
\ee
The fact that the Schr\"odinger potential has a power law rather than exponential falloff near the horizon is characteristic of extremal rather than finite temperature horizons. When a Schr\"odinger equation has an asymptotic region in which the potential has power law falloff, one generically expects that the late time evolution of the wavefunction at some fixed position will be dominated by scattering events in which 
an excitation travels a long distance into the asymptotic region and is reflected back. This occurs at a rate proportional to the asymptotic potential and hence leads to a power law decay in time (see e.g. \cite{Ching:1994bd, Ching:1995tj} for more precise arguments). Fourier transforming to frequency space, the power law tail leads to a branch cut running along the negative imaginary axis from $z=0$. This is the branch cut we are seeing in figure \ref{fig:q0} at $T=0$. In contrast, an exponential falloff near the horizon can be shown to imply that there cannot exist branch cuts emanating from $z=0$ \cite{newton, Hartnoll:2005ju}. The presence of a late time power law tail in extreme Reissner-Nordstr\"om-AdS was previously shown in \cite{Wang:2004bv}, in agreement with the observations we have just made. Furthermore 
\cite{Faulkner:2009wj} have recently exhibited this branch cut by explicitly solving the Schr\"odinger equation near $z=0$.

\subsection{Charged scalars and magnetic field dependence}

\subsubsection{The scales involved: $T, B, \mu$}

We now wish to determine the behavior of the quasinormal modes as a function of the magnetic field
in the low temperature limit. There are three scales characterising the black hole and dual field theory: $T, \mu$ and $B$. The remaining dimensionful quantity is the quasinormal mode frequency $z_\star$.
Because there are no other scales, the underlying strongly coupled theory is a CFT, only the ratios of these dimensionful quantities can be physical. We will implement this freedom as follows.

Firstly, note that the following rescalings completely eliminate $r_+$ and $\gamma$ from our differential equation (\ref{eq:quasinormal})
\be
\hat r = \frac{r}{r_+} \,, \quad \hat z = z r_+ \,, \quad \hat T = T r_+ \,, \quad \hat \mu = \frac{\mu r_+}{\gamma} \,, \quad \hat B = \frac{B r_+^2}{\gamma} \,, \quad \hat q = q \gamma \,.
\ee
The quantities $\{\hat T, \hat \mu, \hat B \}$ are dimensionless and satisfy the constraint, from (\ref{eq:TBM}),
\be
4 \pi \hat T = 3 - \hat \mu^2 - \hat B^2 \,.
\ee
We will use this constraint to eliminate, for instance, $\hat \mu$. We will then find the dimensionless quasinormal modes $\hat z_\star$ as a function of $\hat T$ and $\hat B$ using the algorithm above. Finally, the physical quasinormal frequency in units of the chemical potential is obtained by
\be
\frac{z_\star}{\mu} = \frac{\hat z_\star}{\gamma \hat \mu}  = \frac{1}{\gamma \sqrt{3 - \hat B^2 - 4 \pi \hat T}} \,.
\ee
This will be a quasinormal mode at physical temperature and magnetic field
\be
\frac{T}{\mu} = \frac{\hat T}{\gamma \sqrt{3 - \hat B^2 - 4 \pi \hat T}} \,, \qquad \frac{B}{\mu^2} = \frac{\hat B}{\gamma (3 - \hat B^2 - 4 \pi \hat T)} \,.
\ee

The upshot of the considerations of the previous paragraph will be quasinormal modes of the form
\be
z^{(p)}_\star(\ell) = {\mathcal{F}}^{(p)} \left(\frac{T}{\mu},\frac{B}{\mu^2}, \ell \right) \mu \,,
\ee
for some sequence of functions ${\mathcal{F}}^{(p)}$. Our objective is to explore these functions, particularly the $B$ dependence. This will then allow us to evaluate some of the terms in the sum (\ref{eq:meisterformula}).

\subsubsection{Dependence on the charge $q$ of the scalar field}

Before switching on a magnetic field we make a few observations about the $q$ dependence. Without loss of generality we will focus on positive charges. Taking negative charge $q$ would simply result in a reflection about the imaginary axis:
\be
q \leftrightarrow - q \qquad \Leftrightarrow \qquad z_\star \to - \bar z_\star \,. 
\ee
This follows from taking the complex conjugate of (\ref{eq:M}) together with noting that $z \to - \bar z$ will preserve ingoing boundary conditions at the horizon.

\begin{figure}[ht]
\begin{center}
\includegraphics[height=6.5cm]{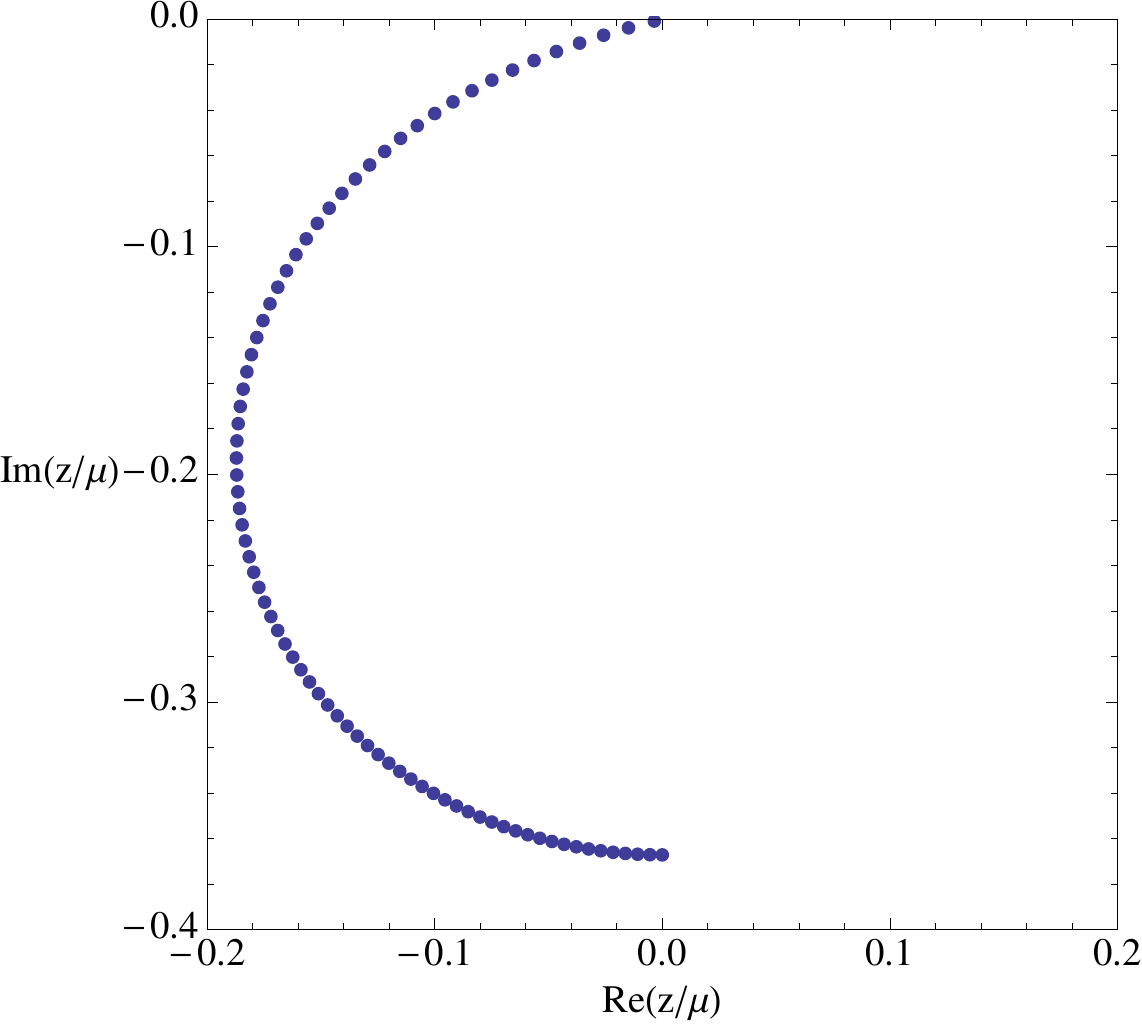}
\end{center}
\caption{The location of the quasinormal mode closest to the real axis as a function of charge. The temperature is $T/\mu = 0.05$ and $B=0$. The charge ranges from $q=0$ to $q = 4.3$. The mass is $m^2=0$ while $\gamma = 1$. At the upper limit of $q$ the mode crosses into the upper half plane, indicating the onset of a superconducting instability.}\label{fig:instability}
\end{figure}

For a given mass squared there exists a critical charge such that if $q$ is larger than the critical value the extremal Reissner-Nordstr\"om black hole becomes unstable. A precise expression for the critical charge as a function of the mass was obtained in \cite{Denef:2009tp},
following the initial discussions of the instability in \cite{Gubser:2008px, Hartnoll:2008vx, Hartnoll:2008kx}.
New instabilities were discovered in (\cite{Faulkner:2009wj}), we briefly discuss these near the end of this section.
The instability indicates the onset of a superconducting phase.
A finite temperature improves the stability of the black holes, although a sufficiently large charge $q$ will always result in an instability. In figure \ref{fig:instability} we show how the location of the quasinormal mode closest to the real axis moves as a function of charge $q$ at a fixed low temperature $T/\mu = 0.05$ and $m^2 = 0$. At $q=0$ the mode is on the imaginary axis, as in figure \ref{fig:q0}. As the charge is increased the mode moves up towards the real axis following an almost semicircular trajectory. At a critical charge $q_c \approx 4.3$ the pole crosses into the upper half plane, indicating the onset of a superconducting instability. This value for $q_c$ agrees nicely with figure 1 in \cite{Denef:2009tp}.

\begin{figure}[h]
\begin{center}
\includegraphics[height=4.5cm]{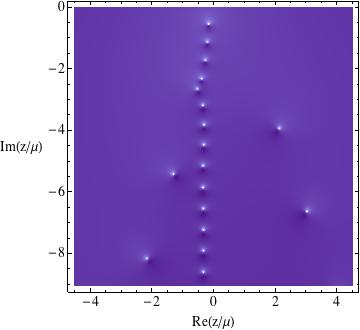}
\includegraphics[height=4.5cm]{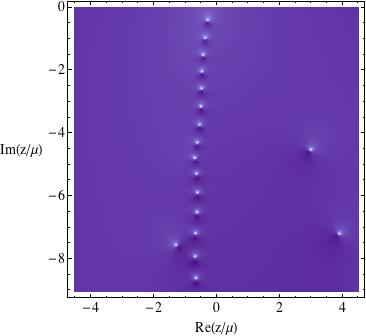}
\includegraphics[height=4.5cm]{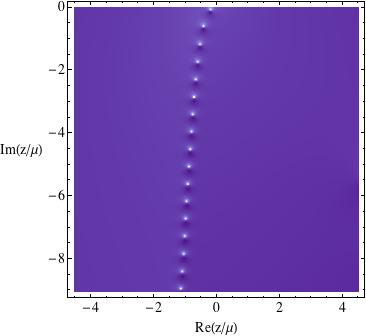}
\end{center}
\caption{Quasinormal modes at temperature $T/\mu = 0.075$. Charges from left to right: $q=1$, $q=2$ and $q=4$. The plots show the lower half frequency plane $z/\mu$, and bright spots denote quasinormal modes. Both plots have $m^2=0$, $\gamma=1$ and no magnetic field.}\label{fig:q1}
\end{figure}

While the onset of superconductivity is the most dramatic effect that occurs as a function of the charge of the scalar field, it is also interesting to see how the higher quasinormal modes rearrange themselves.
The low lying quasinormal modes at low temperature of a scalar field with charges $q=1,2$ and $4$ are shown in figure \ref{fig:q1}. The $q=0$ quasinormal modes for the same mass and temperature were already shown in figure \ref{fig:q0}.

In the plots of figure \ref{fig:q1} we can see how the mode closest to the real axis moves up towards the axis as the charge is increased, as we saw previously in figure \ref{fig:instability}. The motion of the other quasinormal modes is very curious. The line of quasinormal modes that was along the negative imaginary axis at zero charge bends increasingly towards the left. Meanwhile, the modes that were to the left of these in figure \ref{fig:q0} move upwards and to the right, and one by one merge with the sequence of modes that were along the imaginary axis. By the time the charge is $q=4$, the rightmost plot in figure \ref{fig:q1}, it is impossible to distinguish any more between these two types of low lying modes. On the other hand, the modes that were to the right of the imaginary axis are pushed down and further to the right, eventually disappearing from our plot.

Figure \ref{fig:q1} suggests the following interpretation. There are quasinormal modes that for charges less than some critical charge $q < q_\text{crit.}$ remain of order $\mu$ as $T \to 0$, while at sufficiently large charge $q > q_\text{crit.}$ they coalesce with other poles that are forming a branch cut as $T \to 0$. At strictly zero temperature this would presumably correspond to a critical charge at which the pole crosses the branch cut into an `unphysical' sheet. Any given  pole forming the branch tends to the origin as $T \to 0$. Therefore one would have the nonanalytic in $q$ behavior that
\be\label{eq:merging}
\lim_{T \to 0} z_\star = \left\{
\begin{array}{ccc}
{\mathcal{O}}(\mu) & \text{for} & q < q_\text{crit} \\
0 & \text{for} & q > q_\text{crit} 
\end{array}
\right. \,.
\ee
To establish this behavior unambiguously would require higher precision numerics at low temperature than we are currently able to perform.

Varying the charge $q$ corresponds to varying the charge of an operator in the field theory and so is not an operation that can be performed within a given theory. However, we will now see that the merging effect shown in figure \ref{fig:q1} and equation (\ref{eq:merging}) can be undone by a magnetic field. This leads to a possible source of (periodic in $1/B$) nonanalytic behavior of the free energy due to bosons.

\subsubsection{Dependence on $B$: possible periodic nonanalyticities from bosons}

We now turn to the dependence of the quasinormal modes on the magnetic field. We will focus on a case in which, in the absence of a magnetic field, the merging of poles with a branch cut uncovered in figure \ref{fig:q1} has occurred. This requires a sufficiently large charge $q$. As we would like to work at zero temperature, we will increase the mass of the field so that it remains stable at zero temperature. The choice $q=4$, $m^2=10$ (hence $\Delta=5)$ does the job \cite{Denef:2009tp}. We will also work in the limit of large $\ell$ and small $B$, with $B \ell$ order one, so that we can isolate effects that are potentially periodic in $1/B$.

\begin{figure}[h]
\begin{center}
\includegraphics[height=4.5cm]{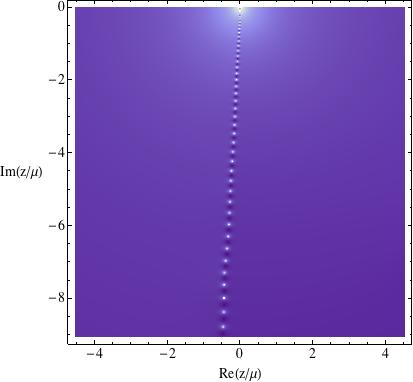}
\includegraphics[height=4.5cm]{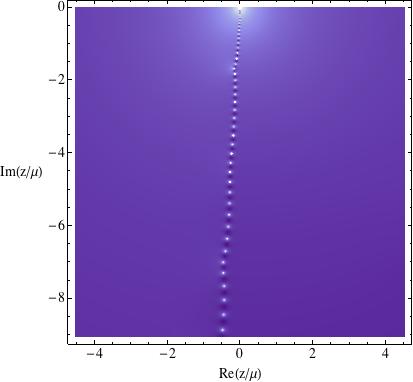}
\includegraphics[height=4.5cm]{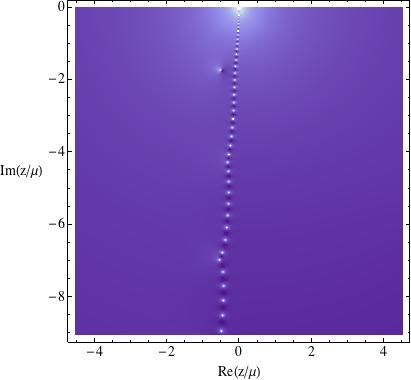} \\
\includegraphics[height=4.5cm]{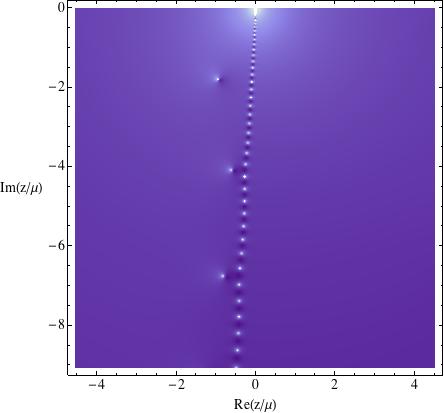}
\includegraphics[height=4.5cm]{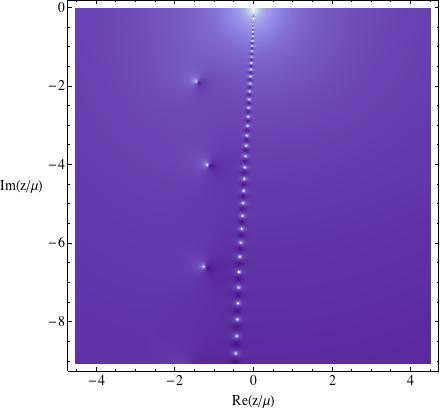}
\includegraphics[height=4.5cm]{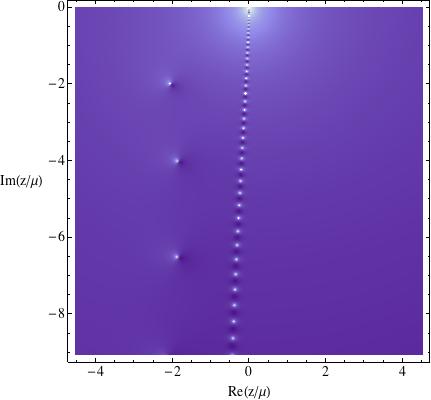}
\end{center}
\caption{Emergence of quasinormal modes from a branch cut as a function of magnetic field and at zero temperature. All plots have $q=4$, $\gamma=1$, $m^2=10$ and $\ell = 10$. The plots show the lower half frequency plane $z/\mu$, and bright spots denote quasinormal modes. From top left $k/\mu \equiv \sqrt{2 \ell |q B|}/\mu \approx 0.283, 0.632, 0.894, 1.26, 1.79, 2.53$. The closely spaced modes are a branch cut that has been discretised into poles by the finite value of $N=500$ in (\ref{eq:Amatrix}).
}\label{fig:Bplots}
\end{figure}

Figure \ref{fig:Bplots} shows the behavior of the $T=0$ quasinormal modes of this charged scalar ($q=4$, $m^2=10$) as a magnetic field is turned on. The top left plot, with a small value of $\ell B$, is analogous to the rightmost plot in figure \ref{fig:q1}. The branch cut bends to the left whereas various poles have either merged with the branch cut or moved off to the right of the plotted region (as in figure \ref{fig:q1}). As the magnetic field is increased, the poles that had merged with the branch cut are seen to reappear, moving out to the left.

Within the current accuracy of our numerics, it is not completely clear whether the poles emerge from the cut at specific finite values of $\ell B$ or whether the pole is distinct from the cut at arbitrarily small $\ell B$. If the former case is true, than this may lead to non-analyticities in the free energy that are periodic in $1/B$, with $q_\text{crit.}$ in (\ref{eq:merging}) replaced by a critical $k_\text{crit.}$ (with $k^2 \equiv 2 \ell |q B|$). Given that such nonanalyticities are usually associated strictly with fermions, it would be very interesting indeed if bosonic operators can lead to these effects at strong coupling. We hope that future work will settle this question.

A possible simple interpretation for why the poles move to the left in figure \ref{fig:Bplots} suggests itself. Recall that in the limit we are working here, the poles in a magnetic field $B$ at some Landau level $\ell$ are the same as the poles in the absence of a magnetic field and at a spatial momentum $k^2 = 2 \ell |q B|$. At zero chemical potential ($\mu = 0$), relativistic invariance will require a branch cut in the $T=0$ retarded Green's function emanating from $k = z$. It may be that the vertically spaced poles in the lower rightmost plots of figure \ref{fig:Bplots} are a finite $\mu$ remnant of this branch cut.

\subsection{Magnetic susceptibility near a superconducting instability}

We have already mentioned that if a bosonic quasinormal mode hits the real axis it typically indicates
the onset of a superconducting instability. We saw an example of this in figure (\ref{fig:instability}) as a function
of the charge of the scalar field. One expects to see a strong response in the susceptibility as the temperature is lowered close to the critical temperature $T_c$, yet this in absent to leading order at large $N$ (i.e. in the classical result of equation (\ref{eq:Omega}) which is unaware of the existence of $T_c$).
In this section we will outline a one loop computation of a divergent magnetic susceptibility. The computation is essentially identical to the standard computation in flat space, see e.g. \cite{schmid}, except that we will use zeta function regularisation. The simple steps below form the conceptual outline of a more complete calculation.

Firstly, we assume that only quasinormal modes close to the origin in frequency space are important. This is reasonable because the divergence we find is directly due to the $\log z_\star$ divergence of our formula (\ref{eq:meisterformula}) as $z_\star \to 0$. At least one such mode exists, by assumption, because $T \sim T_c$. To obtain the correct susceptibility we must sum over the higher Landau levels associated to this mode \cite{schmid}. These are also close to the origin in the limit of small magnetic field, a limit we will assume in this calculation. In fact, the computation we are about to present will require
\be
B \ll (T - T_c) T_c \,.
\ee
However, we will not sum over the `excited' quasinormal modes, as these will be at a distance at least of order $T_c$ from the origin. Using the fact that $\Gamma(x) \sim 1/x + \cdots$, as $x \to 0$, then for the modes of interest, $z_\star \sim 0$, (\ref{eq:meisterformula}) reduces to
\be\label{eq:smallzsum}
\Omega_B = \frac{|q B| A T}{2 \pi} \sum_\ell \text{Re} \, \log \frac{z_\star(\ell)}{T} + \cdots \,.
\ee

Secondly, we assume that the quasinormal modes appearing in (\ref{eq:smallzsum}) can be written as
\be
z_\star(\ell) = \a_1  |q B| (\ell + \half) + \a_2 (T -T _c) + \cdots \,,
\ee
with $\a_i$ complex constants with negative imaginary parts. The temperature dependence is fixed by the requirement that $z_\star$ go to zero at $T= T_c$ in the absence of a magnetic field. The magnetic field and Landau level dependence at a linearised level at small $B$ is fixed by the fact that only the combination $B (\ell + \half)$ appears in the differential operator (\ref{eq:M}). The offset of $\half$ is crucial in this computation.
Thus the free energy (\ref{eq:smallzsum}) becomes, using zeta function regularisation,
\be
\Omega_B = - \frac{A T}{2 \pi} \text{Re} \, \zeta_B'(0) \,,
\ee
with
\be
\zeta_B(s) = \sum_\ell \frac{|q B|}{\left(\a_1 |q B| (\ell + \half) + \a_2 (T -T _c)  \right)^s} = |q B| (\a_1 |q B|)^{-s} \zeta_H \left(s, \frac{1}{2} + \frac{\a_2 (T-T_c)}{\a_1 |q B|} \right) \,,
\ee
where as previously $\zeta_H(s,x)$ is the Hurwitz zeta function.

To leading order in $B \ll (T - T_c) T_c$ this gives a divergent susceptibility
\be
\chi = - \frac{\pa^2 \Omega_B}{\pa B^2} = - \frac{A T}{24 \pi}  \text{Re} \left( \frac{\a_1}{\a_2} \right) \frac{1}{T - T_c} \,.
\ee
If the real part of $\a_1/\a_2$ is positive, then
the divergence is diamagnetic, as we should expect for the onset of superconductivity.
Numerical computations suggest that this is the case.
As previously, the divergence indicates the breakdown of perturbation theory and the need to resum higher order corrections. This divergence is a second vivid example of physics beyond the classical gravity limit, albeit not the focus of the present paper. It is of interest to flesh out the computation outlined above.

A different order of limits is to take $T=0$ with a large magnetic field stabilising the superconducting instability.
One then lowers the magnetic field, finding an instability at $B = B_{c2}$ \cite{Albash:2008eh, Hartnoll:2008kx}. Distinct to the case we have just treated, this would be a zero temperature quantum phase transition. Furthermore the zero temperature ground state for $B < B_{c2}$ is not known in this regime.  It is plausible that we can use the same logic as in sections \ref{sec:nonanalytic} -- \ref{sec:zeroT}. The relevant (likely $\ell=0$) quasinormal mode should behave for $B$ just above $B_{c2}$ like
\be
\text{Im} \, z_\star(0) \sim - |B - B_{c2}|^\delta \,,
\ee
for some positive $\delta > 0$. The bosonic analogue of (\ref{eq:chistrong}) will then give
\be
\chi \sim \pm |q B| A  \, |B - B_{c2}|^{\delta-2} \,.
\ee
Interestingly, the sign depends on the value of the angle $\theta$ of section \ref{sec:zeroT} which now lies in the range $\pi (1 - 2\nu) < \theta < \pi$. For cases in which the divergence is paramagnetic rather than diamagnetic, it may be that the state at $B < B_{c2}$ is not superconducting at $T=0$. To fully address this question one should generalise the results of \cite{Faulkner:2009wj} for scalars in a magnetic field. We cannot simply use the existing results as these translate into statements about large $\ell$ modes, while the $T=0$ finite $B$ instability is likely sensitive to the lowest Landau level only, with the higher levels being `gapped' by the magnetic field.

\subsection{WKB results for charged scalar quasinormal modes}

As a check on our numerics, and also with a view to eventually controlling the UV behavior of the sums over quasnormal modes in (\ref{eq:meisterformula}) and (\ref{eq:meisterfermions}), we have computed the large frequency limit of the charged bosonic quasinormal modes, at fixed $\ell$. The derivation is given in appendix \ref{sec:wkb}, closely following the WKB analysis of neutral massless scalars in the AdS-Reissner-Nordstrom background in \cite{Natario:2004jd}. The method used for obtaining the asymptotic behavior of quasinormal modes was pioneered by \cite{Motl:2003cd}.

The result is that the asymptotic quasinormal modes for the equation (\ref{eq:M}) are
\begin{equation}
 \int_0^{x_0} dx \left( z_\star - q \mu \frac{(r_+-r(x))}{r_+}  \right)  = n \pi + \frac{1}{2i} \log \left( 2 \cos \left(\frac{\pi}{6} \right) \right) +  \frac{\pi( 2 + \sqrt{4 m^2 + 9})}{4} \,, \label{eq:WKB}
 \end{equation}
with $n \in \N$, while $x_0$ and $x(r)$ are given by
\be
x_0 = \sum_{p=1}^{4} \frac{1}{f'(r_p)} \log \left(- \frac{r_+}{r_p} \right) \,, \qquad  x(r) = - \sum_{p=1}^{4} \frac{1}{f'(r_p)} \log \left(1- \frac{r_p}{r} \right) \,,
\ee
where $r_p$ are the zeroes of $f(r)$. Thus $r(x)$ in (\ref{eq:WKB}) must be found by inverting the second of these expressions. The integral in (\ref{eq:WKB}) is along the Stokes line specified in \cite{Natario:2004jd}. In the complex $x$ plane this is simply a straight line connecting $0$ and $x_0$. In computing $r(x)$ one must be careful to remain on the correct sheet. The `mirror' poles, $z_\star \to - \bar z_\star$ and $q \to - q$, are obtained with a different choice of Stokes lines.

The result (\ref{eq:WKB}) does not depend on the Landau level $\ell$ and so only depends on $B$ through $f(r)$. To capture the Landau level dependence will require a more refined analysis, perhaps along the lines of \cite{Festuccia:2008zx}. The complication is that the large $\ell$ and large $n$ limits do not commute. Ideally we would like a single formula encapsulating both of these limits.

We have checked that (\ref{eq:WKB}) agrees with our numerical results for the sequence of poles extending down in the complex plane with a spacing set by $\mu$ (the slope agrees very well, there is a slight mismatch in the offset which we believe is due to not being able to access the large frequency limit with our numerics). A curious feature, however, is that the sequence of poles coalescing to form a branch cut, i.e. those that appear more closely spaced at low temperatures in figures (\ref{fig:q0}) and (\ref{fig:q1}), do not appear in this formula. This may suggest that the zero temperature branch cut emanating from the origin terminates at some large but finite frequency rather than extending asymptotically. We hope this can be determined precisely in future work.

\section{Discussion}

In this paper we have argued that one loop effects in the bulk contain information that
needs to be accessed if the `applied holography' research program is to successfully disentangle the
different physical contributions to observables such as the free energy and electrical conductivity.
We illustrated this point of view by showing that the one loop contribution to the magnetic susceptibility exhibits
the de Haas-van Alphen quantum oscillations that are not manifest at a bulk classical level.
In achieving this, we discovered that the oscillations at strong coupling differ from their weak
coupling counterparts in that the periodic low temperature nonanalyticities of the magnetic
susceptibility are not delta functions but rather power law divergences.

The key step in our arguments has been the (new to our knowledge) observation that one loop determinants can be expressed as a sum over quasinormal modes. The quasinormal frequencies give the poles of
retarded Green's functions in the strongly coupled dual field theory
and are therefore natural physical quantities. In particular,
if one of these poles comes close to the real frequency axis then we may be able to think of it as
an emergent quasiparticle excitation
of sorts (although it may not have a finite residue \cite{Faulkner:2009wj}). It may therefore be
sensible to think about the isolated individual contribution of this pole to physical quantities. In our computation
it was precisely such poles near the real axis which lead to the de Haas-van Alphen oscillations. More generally
it might be interesting to compute the one loop contribution of such poles to other
quantities such as the electrical conductivity.

We primarily examined the low temperature regime in this paper, as the sum over quasinormal
modes simplified in this limit and the nonanalytic motion of a specific pole contained the physics
of interest. In general one would like to exhibit quantum oscillations at the higher temperatures
more relevant for comparison with recent experiments. In particular, it would be interesting to
obtain a strong coupling analogue of the Lifshitz-Kosevich formula \cite{shoenberg}. To deal with higher temperatures
it will likely be necessary to Poisson resum the Landau levels. Before this can be done,
the asymptotic behavior of the quasinormal poles will need to be well characterised. We took some
first steps in this direction above by computing the WKB form of the quasinormal poles at a fixed
Landau level $\ell$. It would also be of interest to study the charged fermionic quasinormal modes
numerically in a similar way to our studies of charged bosons in this paper.

At the low temperatures we have studied, quantum oscillations are usually closely connected to the
quantum Hall effect. It seems possible that the novel nonanalytic behavior of the low temperature
free energy as a function of the magnetic field in theories with gravitational duals may indicate
nonconventional quantum Hall states.

In our numerical studies of the quasinormal frequencies of charged bosons we uncovered some nontrivial
dependence of the pole locations on the charge of the field and the background magnetic field. It would be very interesting to improve the numerical determination of the positions of the poles, perhaps by implementing a better numerical algorithm than the rather general method used in this paper. A more accurate knowledge of the `pole dancing'
will be necessary to establish unambiguously whether there are nonanalyticities as a function of the magnetic field due to bosonic quasinormal modes `disappearing' into branch cuts. Also, it would be useful to be able to
extend the numerics further into the lower half frequency plane and determine whether the branch cut terminates or whether it continues asymptotically.

\section*{Acknowledgements}

All the authors are pleased to acknowledge the hospitality of the KITP in Santa Barbara while this work was being completed. Over the course of this work we have enjoyed stimulating discussions with Lars Fritz, Matt Headrick, Diego Hofman, Sung-Sik Lee, Hong Liu, John McGreevy, Andy Neitzke and Rafael Porto. 
This research was supported by the National Science Foundation under grant DMR-0757145 (SS), by the FQXi
foundation (SAH and SS), by a MURI grant from AFOSR (SS), by DOE
grant DE-FG02-91ER40654 (FD and SAH). The research at the KITP was supported in 
part by the National Science Foundation under Grant No. PHY05-51164.

\appendix

\section{Schr\"odinger form}

The equation (\ref{eq:MBeval}) can be converted into a Schr\"odinger form (albeit with a complex potential at general $z$)
\be
- \frac{d^2 \Psi}{d r_\star^2} + V(r_\star) \Psi = L^2 \lambda \Psi \,,
\ee
by setting
\be\label{eq:phimap}
\phi = \frac{r^{3/2}}{f^{1/4}} \Psi \,, \qquad dr_\star = \frac{dr}{r f^{1/2}} \,.
\ee
This shows that the required norm is
\be
\int d r_\star |\Psi|^2 = \int \frac{dr}{r^4} |\phi|^2 < \infty \,.
\ee
The potential is, written in terms of $r$,
\be
V = L^2 m^2 + K_\ell r^2 - \frac{r^2}{f}\left(z - q\mu \left(1 - \frac{r}{r_+}\right) \right)^2 + \frac{9 f}{4} + \frac{r^2 f''}{4} - \frac{5 r f'}{4} - \frac{r^2 f'^2}{16 f} \,.
\ee
At the boundary at infinity, $r=0$, the potential tends to $V(0) = \frac{9}{4} + (L m)^2$ which is positive so long as the scalar field satisfied the Breitenlohner-Freedman bound. This naturally suggests a continuum of eigenvalues at $\lambda \geq \frac{9}{4L^2} + m^2$.

For self-adjoint operators one can prove that a complete basis of eigenstates lives in an extended Hilbert space which allows certain non-normalisable functions. 
At vanishing charge ($q=0$) and pure imaginary frequencies
 $z=i \w_n$ the operator $M(z,\ell)$ is self-adjoint and positive. From the definition of the extended Hilbert space  (see e.g. \cite{ballentine}) one can easily check that in this case there is a basis of eigenfunctions for which the condition $\lambda \geq \frac{9}{4L^2} + m^2$ holds. While the operator is not self-adjoint on the imaginary $z$ axis with nonzero charge and chemical potential, these do not alter the boundary behavior of the field and we expect the continuum to persist as a basis of states. This statement is routinely used for charged fields in flat space and finite chemical potential. The only generalisation we are making is to consider a curved spacetime background.

\section{Damped harmonic oscillator}
\label{sec:damped}

In this appendix we show how some of the methods we have developed in this paper can be applied to the very simple case of a damped harmonic oscillator.

Consider the retarded Green's function 
\begin{equation}
G^R (z) =  \frac{1} {- (z-\mu)^2 -  2 i \gamma z + m^2} \,,
\label{f0}
\end{equation}
with $|\mu| < m$ and $\gamma > 0$. This is just the response function of a damped simple harmonic
oscillator. All the poles of $G^R (z)$ are in the lower half frequency plane.
Associated with this retarded Green's function, we have the Matsubara Green's function 
defined on the imaginary frequency axis by
\begin{equation}
G (i \omega_n) = \frac{1}{(\omega_n + i \mu)^2 + 2 \gamma |\omega_n| + m^2} \,.
\label{e0}
\end{equation}

We now use the spectral representation to write
\begin{equation}
G(z) = \int_{-\infty}^{\infty} \frac{d \Omega}{\pi}  \mbox{Im} \left[ \frac{1} {- (\Omega-\mu)^2 -  2 i \gamma \Omega + m^2} \right] \frac{1}{(\Omega - z)} \,.
\label{e2}
\end{equation}
Note that $G(z) = G^R (z)$ in the upper half plane, but $G (z) \neq G^R (z)$ in the lower half plane. However, on the imaginary frequency axis, with $z= i \omega_n$, the expressions in Eq. (\ref{e2}) and (\ref{e0}) agree for all positive and negative $\omega_n$. Note that (\ref{e2}) defines an analytic function $G(z)$ which has a branch-cut
on the real $z$ axis, and no poles anywhere.

We can use the spectral representation to perform the frequency summation
\begin{equation}
T \sum_{\omega_n} G (i \omega_n) = \int_{-\infty}^{\infty} \frac{d \Omega}{\pi}  \mbox{Im} \left[ \frac{1} {- (\Omega-\mu)^2 - 2 i \gamma \Omega + m^2} \right] \frac{1}{e^{\Omega/T} - 1} \,.
\label{e3}
\end{equation}
This object is of interest because it is related to the free energy ${\mathcal{F}}$ by
\be\label{eq:dfdm}
\frac{d{\mathcal{F}}}{dm^2} =  T \sum_{\omega_n} G (i \omega_n)  \,.
\ee
We can try to simplify (\ref{e3}) by using the position of the poles of $G^R$:
\begin{equation}
G^R (z)  = \sum_i \frac{c_i}{z_i - z} \,,
 \label{e3a}
\end{equation}
with $\mbox{Im} [z_i] < 0$. Explicitly the poles are $z_{1,2}$ and they obey
\begin{eqnarray}
z_1 + z_2 &=& 2 \mu - 2 i \gamma \,, \nonumber \\
z_1 z_2 &=& \mu^2  - m^2 \,, \nonumber \\
z_{1,2} &=& \mu - i \gamma \pm \sqrt{m^2 - \gamma^2 - 2 i \gamma \mu} \,. \label{zres}
\end{eqnarray}
Note that at large $z$ the left hand side of (\ref{e3a}) goes like $\sim - 1/z^2$ which implies from the right hand side that 
\begin{equation}
\sum_i c_i = 0 \,.
\label{c1}
\end{equation}
This condition may often hold more generally independently of the number of poles, and was implicitly used in performing the frequency summation in Eq.~(\ref{e3}). So we have
\begin{equation}
T \sum_{\omega_n} G (i \omega_n) = \sum_i \int_{-\infty}^{\infty} \frac{d \Omega}{\pi} \mbox{Im} \left[
\frac{c_i}{z_i - \Omega} \right] \frac{1}{e^{\Omega/T} - 1} \,.
\label{e4}
\end{equation}
Convergence of the integral at $\Omega=0$ requires that 
\begin{equation}
\sum_i \mbox{Im} \left[
\frac{c_i}{z_i} \right] = 0,
\label{e4a}
\end{equation}
which is also seen to hold from Eqs.~(\ref{e3a}) and (\ref{f0}). Again, we expect Eq.~(\ref{e4a}) to hold more generally.
It does not seem to be possible to simplify the integral in Eq.~(\ref{e4}) further, in general.
However, at $T=0$, we obtain
\begin{equation}
T \sum_{\omega_n} G (i \omega_n) = \frac{1}{\pi} \sum_i \mbox{Im} \Bigl[ c_i \log (z_i) \Bigr].
\end{equation}

Now we turn to the free energy. Let us write
\begin{equation}
[G^R (z)]^{-1} = f(z) + m^2 \,,
\end{equation}
Then, near a pole $z_i$, 
\begin{equation}
G^R (z) \approx - \frac{1}{f' (z_i)} \frac{1}{(z_i -z)} \,.
\end{equation}
But $f(z_i) = - m^2$ and so 
\begin{equation}
f'(z_i) \frac{d z_i}{d m^2} = -1 \,,
\end{equation}
and hence
\begin{equation}
G^R (z) \approx \frac{dz_i}{d m^2} \frac{1}{(z_i -z)} \,.
\end{equation}
So we have
\begin{equation}
c_i = \frac{dz_i}{dm^2}\,.
\end{equation}
This is an interesting result linking the residue and location of the `quasinormal poles'.
With this expression, we can write the constraints in Eq.~(\ref{c1}) and (\ref{e4a}) as
\begin{eqnarray}
\frac{d}{dm^2} \sum_i z_i &=& 0 \,, \nonumber \\
\frac{d}{dm^2} \sum_i \mbox{Im} \Bigl[ \log z_i \Bigr] &=& 0 \,.
\end{eqnarray}
For the frequency summation, we have
\begin{equation}
T \sum_{\omega_n} G (i \omega_n) = \sum_i \int_{-\infty}^{\infty} \frac{d \Omega}{\pi}  \frac{d}{d m^2} \mbox{Im} \left[
\log(z_i - \Omega) \right] \frac{1}{e^{\Omega/T} - 1} \,.
\label{e40}
\end{equation}
Integrating this, using (\ref{eq:dfdm}), we obtain
\begin{equation}
\mathcal{F} = \sum_i \int_{-\infty}^{\infty} \frac{d \Omega}{\pi}  \mbox{Im} \left[
\log(z_i - \Omega) \right] \frac{1}{e^{\Omega/T} - 1} \,,
\end{equation}
up to $m$-independent terms. This expression is analogous to (\ref{eq:OmegaBH}) in the main text.

For the damped harmonic oscillator, we obtain as $T \rightarrow 0$:
\begin{eqnarray}
\left. \mathcal{F}\right|_{T \to 0} &=& - \sum_i \int_0^\infty \frac{d \Omega}{\pi} \mbox{Im} \left[\log (z_i + \Omega) \right] - \frac{4}{\pi} \frac{\gamma}{(m^2 - \mu^2)} \int_0^{\infty} \frac{ \Omega d \Omega}{e^{\Omega/T} - 1} \nonumber \\
&=&  \frac{1}{\pi} \sum_i \mbox{Im} \left[ z_i \log z_i \right]  - \frac{2\pi}{3} \frac{\gamma T^2}{(m^2 - \mu^2)}  \,,
\label{tsq}
\end{eqnarray}
where the $z_i$ are given in Eq.~(\ref{zres}).

Alternatively, we can write
\begin{equation}
T \sum_{\omega_n} G (\omega_n) = \sum_i \left[ T \sum_{\omega_n > 0} \frac{c_i}{z_i - i \omega_n} + T \sum_{\omega_n \leq 0} \frac{c_i^\ast}{z_i^\ast - i \omega_n} \right] \,.
\end{equation}
Integrating this with respect to $m^2$ we obtain
\begin{equation}
\mathcal{F} =  \sum_i \left[ T \sum_{\omega_n > 0} \log (z_i - i \omega_n )+ T \sum_{\omega_n \leq 0} \log (z_i^\ast - i \omega_n)  \right] \,. \label{g1}
\end{equation}
This formula is analogous to (\ref{eq:OmegaBH2}) in the main text
We can now evaluate the frequency summation by using the identities of zeta function regularization in equation (\ref{eq:zetareg}). This yields
\begin{eqnarray}
\mathcal{F} &=&  \sum_i \left[ T \sum_{n=1}^\infty \log \left( n + \frac{iz_i}{2 \pi T} \right)+ 
T \sum_{n=0}^{-\infty} \log \left(n + \frac{iz^\ast_i}{2 \pi T} \right)  \right] \nonumber \\
 &=&  \sum_i \left[ T \sum_{n=0}^\infty \log \left( n + \frac{iz_i}{2 \pi T} \right)+ 
T \sum_{n=0}^{\infty} \log \left(n - \frac{iz^\ast_i}{2 \pi T} \right) - T \log \left[ \frac{|z_i|}{2 \pi T} \right] \right] \nonumber \\
&=& - T \sum_i \log \left[ \frac{  |z_i | \left| \Gamma \left( i z_i /(2 \pi T) \right) \right|^2}{4 \pi^2 T} \right] \,. \label{y2}
\end{eqnarray}
This result is analogous to (\ref{eq:meisterformula}) in the main text.

As $T \rightarrow 0$ at fixed $z_i$, we can use the asymptotic expansion of the $\Gamma$ function
\begin{equation}
\log \Gamma (x) \sim \left(x - \frac{1}{2} \right) \log x - x + \frac{1}{2} \log (2 \pi) + \frac{1}{12 x} \,. \label{Gam}
\end{equation}
Working with all the terms in Eq.~(\ref{Gam}), we have the complete expression as $T \rightarrow 0$, assuming all the $z_i$ remain finite in this limit
\begin{eqnarray}
\left.  \mathcal{F} \right|_{T \to 0}  &=& \sum_i \Big[ - T \log \left( \frac{ |z_i|}{4 \pi^2 T} \right)  + \frac{\mbox{Im}(z_i)}{\pi} \log \left( \frac{1}{2 \pi  e T} \right) \nonumber \\
& + & \frac{1}{\pi} \mbox{Im} \left[ z_i \log (i z_i) \Big]
+ \frac{T}{2} \log \left( \frac{|z_i|^2}{4 \pi^2 T^2} \right) - T \log (2 \pi) - \frac{T}{6} \mbox{Im} \left( \frac{2 \pi T}{z_i} \right) \right] + \mathcal{O} (T^3)  \nonumber \\
&=& \frac{\log(1/(2 \pi e T))}{\pi} \sum_i \mbox{Im}(z_i)  + \frac{1}{2} \sum_i \mbox{Re} (z_i) \nonumber \\
& + & \frac{1}{\pi} \sum_i \mbox{Im} \left[ z_i \log z_i \right] - \frac{\pi T^2 }{3} \sum_i \mbox{Im} \left( \frac{1}{z_i} \right)  + \mathcal{O}(T^3) .
\end{eqnarray}
The above expression is very general. Specializing to the values in Eq.~(\ref{zres}) we obtain
\begin{eqnarray}
\left.  \mathcal{F} \right|_{T \to 0}  
&=& - \frac{2 \gamma}{\pi} \log \left( \frac{1}{2 \pi e T} \right) + \mu + \frac{1}{\pi} \sum_i \mbox{Im} \left[ z_i \log z_i \right] - \frac{2 \pi }{3} \frac{\gamma T^2}{(m^2 - \mu^2)} + \mathcal{O}(T^3) \,.
\end{eqnarray}
Apart from the first two $m$-independent terms, this matches precisely with Eq.~(\ref{tsq}).
 
\section{Asymptotic frequencies}
\label{sec:wkb}

In this appendix we outline the derivation of the WKB formula (\ref{eq:WKB}) for the charge quasinormal modes in the main text. We will follow the notation of \cite{Natario:2004jd} which uses different coordinates to those in this paper. Let us therefore rewrite our equation (\ref{eq:M}) as
\begin{equation}
\left[ - R^4 \frac{d}{d R} \left( \frac{F}{R^2} \frac{d}{dR} \right) - \frac{R^2}{F} \left(
\omega - q \mu \left( 1 - \frac{R}{r_+} \right) \right)^2 + ( K_\ell R^2 + m^2 ) \right] \Phi (R) = 0 \,, \label{eq:rewrite}
\end{equation}
where
\begin{equation}
F(R) = 1 - 4 \left(\frac{R}{r_+} \right)^3 ( 1 - \pi r_+ T) +  \left(\frac{R}{r_+} \right)^4 ( 3 - 4 \pi r_+ T) \,.
\end{equation}
The boundary is at $R=0$ and the horizon is at $R=r_+$. We are interested in the quasinormal
frequencies $\w$ of this equation.

We want to transform equation (\ref{eq:rewrite}) into the notation of Ref.~\cite{Natario:2004jd}.
So we define
\begin{eqnarray}
r &=& \frac{r_+}{R} \,, \nonumber \\
f(r) &=& r_+ R^{-2} F(R) \nonumber \\
&=& \frac{r^2}{r_+} - \frac{4 ( 1 - \pi r_+ T)}{r r_+} +  \frac{( 3 - 4\pi r_+ T)}{r^2 r_+} \,, \nonumber \\
\phi (r) & = & r \Phi (R) \,.
\end{eqnarray}
Now the boundary is at $r = \infty$ and the horizon is at $r=1$, and the differential equation
transforms to
\begin{equation}
-  f \frac{d}{dr} \left(  f \frac{d \phi}{d r} \right)  + \left( \frac{m^2}{r_+} + \frac{K_\ell r_+}{r^2} + 
\frac{1}{r} \frac{d f}{dr} \right) f \phi = \left(\omega - q \mu \frac{(r - 1)}{r} \right)^2 \phi \,. \label{e1}
\end{equation}
This is similar to Eq.~(3.2) of Ref.~\cite{Natario:2004jd}, allowing us to map some of their results to ours.

We will need the roots of $f(r)$ below. 
The roots are at $r=r_p$ with
\begin{eqnarray}
r_1 &=& 1~~;~~f'(r_1) = 4 \pi T \,,  \nonumber \\
r_2 &=& r_- /r_+ ~~~;~~~0 < r_2 < 1~~~;~~~f'(r_2) < 0 \,, \nonumber \\
r_{3,4} &=& - (1+ r_-/r_+)/2 \pm i \sigma \,.
\end{eqnarray} 
where $\sigma$ is real. These roots satisfy some useful identities:
\begin{eqnarray}
\sum_{p=1}^4 r_p &=& 0 \,, \nonumber \\
\sum_{p=1}^4 \frac{r_p^n}{f'(r_p)} &=& 0~\mbox{for $n=2,0,-1,-2$} \,, \nonumber \\
\sum_{p=1}^4 \frac{r_p}{f'(r_p)} &=& r_+ \,.
\label{iden}
\end{eqnarray}

The WKB formulae will involve contour integrals which are most easily expressed in terms of
a coordinate $x$, related to $r$ by
\begin{equation}
dx = \frac{dr}{f(r)} \,.
\end{equation}
This can be integrated to obtain
\begin{equation}
x (r) = \sum_{p=1}^{4} \frac{1}{f'(r_p)} \log \left( 1 - \frac{r}{r_p} \right) \,.
\label{xr}
\end{equation}

The WKB limit of the quasinormal modes is captured by matching together the behavior of solutions to
(\ref{e1}) near several special values of $r$. We now consider these values in turn.

\subsection*{$r=0$ (black hole singularity)}

The branch-cuts in Eq.~(\ref{xr}) should be chosen so that there is no monodromy around
$r=0$, and also none around $r=\infty$. Then the expansion near
$r=0$ is
\begin{equation}
x = \frac{r_+}{3(3 - 4 \pi r_+ T)} r^3 + \ldots \,,
\end{equation}
and the differential equation (\ref{e1}) becomes
\begin{equation}
- \frac{d^2 \phi}{d x^2} - \frac{2}{9 x^2} \phi = \omega^2 \phi . \label{r0}
\end{equation}
This has the same form as that in Section 3.3.2 of Ref.~\cite{Natario:2004jd} with 
\begin{equation}
j=1/3.
\end{equation}
Note that for $|x| \ll 1$, the term proportional to $q$ in Eq.~(\ref{e1}) is subdominant to either
the $\omega^2$ or the $1/x^2$ terms in Eq.~(\ref{r0}). This means that we can solve
Eq.~(\ref{r0}) in terms of Bessel functions. In the regime where $\omega x \gg 1$, but
with $|x| \ll 1$, we have the solution as in Eq.~(3.40) of \cite{Natario:2004jd}:
\begin{equation}
\phi (x) \sim 2 B_+ \cos\left( \omega x - \alpha_+ \right)  + 2 B_- \cos \left(\omega x - \alpha_- \right)  \,,\label{r0a}
\end{equation}
where $\alpha_+ = \pi (1\pm j)/4$.

However, for the matching below, we have to extend this solution to $x \approx 1$. In this regime
we can use the WKB method, where the right hand side of Eq.~(\ref{e1}) dominates over the potential
on the left hand side. From this method, with $|\omega| \gg 1$, we find that we must replace Eq.~(\ref{r0a}) by
\begin{equation}
\phi (x) \sim 2 B_+ \cos\left( \int_0^x dx \left( \omega - q \mu \frac{(r-1)}{r}  \right) - \alpha_+  \right)  + 2 B_- \cos \left( \int_0^x dx \left( \omega - q \mu \frac{(r-1)}{r}  \right)  - \alpha_-  \right) \label{r0b}
\end{equation}
The integral is to be taken over a suitable contour in the complex plane, which we will discuss below.

\subsection*{$r = \infty$ (asymptotic AdS region)}

Here we have
\begin{eqnarray}
x_0 &=& x ( r \rightarrow \infty ) \nonumber \\
&=&  \sum_{p=1}^{4} \frac{1}{f'(r_p)} \log \left(- \frac{1}{r_p} \right), \label{xo}
\end{eqnarray}
where $x_0$ is a complex number which will play an important role below. Despite the ambiguity in the branch cuts of the logarithms,  
the value of $x_0$ is unique: the identities in Eq.~(\ref{iden}) ensure that there is no monodromy
around $r=\infty$ provided none of the branch cuts extend to $r=\infty$. For $x$ close to $x_0$, we have
\begin{equation}
r \approx \frac{r_+}{x_0 - x},
\end{equation}
and the differential equation Eq.~(\ref{e1}) becomes
\begin{equation}
- \frac{d^2 \phi}{d x^2} + \frac{m^2 + 2}{(x_0 - x)^2} \phi = \omega^2 \phi \,. \label{ei}
\end{equation}
Again, the $q\mu$ term is subdominant everywhere for $|x-x_0| \ll 1$.
In the limit of large $\omega$ this is the same as in Section 3.3.2 of 
Ref.~\cite{Natario:2004jd} with
\begin{equation}
j_\infty = \sqrt{4m^2 + 9}. 
\end{equation}
As $x \rightarrow x_0$, the solutions
of Eq.~(\ref{ei}) are of the form
\begin{equation}
\phi \sim r^{(-1 \pm \sqrt{4m^2 + 9})/2} \,.
\end{equation}
implying that in terms of our original variables $\Phi \sim R^{(3 \pm \sqrt{4m^2 + 9})/2},$ as expected.

The connection to the regime where $|x-x_0| \ll 1$, but $\omega (x_0 - x) \gg 1$,
can be performed using Bessel functions, as in Ref.~\cite{Natario:2004jd}.
Requiring falloff at infinity yields the result above (3.41) in \cite{Natario:2004jd}:
\begin{equation}
\phi (x) \sim 2 C_+ \cos (\omega (x-x_0) + \beta_+) \,,
\label{ria}
\end{equation}
with $\beta_+ = \pi(1+j_\infty)/4$.
As for Eq.~(\ref{r0a}), we have to extend this result for $x_0 - x \sim 1$, but $|\omega| \gg 1$.
Here the WKB method shows that Eq.~(\ref{ria}) is replaced by
\begin{equation}
\phi (x) \sim 2 C_+ \cos \left(   \int_{x_0}^x dx \left( \omega - q \mu \frac{(r-1)}{r}  \right) + \beta_+ \right) \,,
\label{rib}
\end{equation}
where the contour of integration is left unspecified for now.

\subsection*{$r=1$ (black hole horizon)}\

Near the horizon of the black hole
\begin{equation}
x \approx \frac{\log (r-1)}{4 \pi T} \rightarrow -\infty \,.
\end{equation}
The equation (\ref{e1}) becomes
\begin{equation}
- \frac{d^2 \phi}{d x^2} = \omega^2 \phi \,,
\end{equation}
and so the solution is 
\begin{equation}
\phi (r) \approx (r-1)^{\pm i \omega/(4 \pi T)} \,,
\end{equation}
which is as expected. We will want to impose ingoing boundary conditions, as in the main text.

\subsection*{Asymptotic frequencies}

We now see that the determination of the frequencies of the quasi-normal modes 
can be mapped onto the solution in Section 3.3.2 of Ref.~\cite{Natario:2004jd},
after using the values of $j$, $j_\infty$ and $x_0$ quoted above, and including  the phase
shifts from the $q \mu$ term.

Matching the behavior near $r=0$ in Eq.~(\ref{r0b}), with that near $r=\infty$ in Eq.~(\ref{rib}),
we find that the condition (3.41) in \cite{Natario:2004jd} is replaced by
\begin{equation}
\exp\left\{ 2 i  \left(  \int_0^{x_0} dx \left( \omega - q \mu \frac{(r-1)}{r}  \right)  - \beta_+  \right) \right\} 
= \frac{B_+ e^{i \alpha_+ } + B_- e^{i \alpha_-}}{B_+ e^{-i \alpha_+ } + B_- e^{-i \alpha_-}} \,,
\end{equation}
where the integral is to be taken along contour B connecting these points
in Fig.~15 of Ref.~\cite{Natario:2004jd}. Namely, it is the Stokes line satisfying $\text{Im}\, \w x = 0$.
This is crucial in order to be able to accurately distinguish ingoing and outgoing modes at the
horizon.

The matching between $r=0$ and the horizon $r=1$ remains as  in Ref.~\cite{Natario:2004jd}, and we 
have the result above their (3.42).
\begin{equation}
B_+ e^{-3 i \alpha_+} + B_- e^{-3 i \alpha_-} = 0.
\end{equation}

The final result for $\omega$ is obtained by solving these last two equations,
and this shows that Eq.~(3.42) of Ref.~\cite{Natario:2004jd} is replaced by
\begin{equation}
 \int_0^{x_0} dx \left( \omega - q \mu \frac{(r-1)}{r}  \right)  = n \pi + \frac{1}{2i} \log \left( 2 \cos \left(\frac{\pi}{6} \right) \right) +  \frac{\pi( 2 + \sqrt{4 m^2 + 9})}{4} \,,
 \end{equation}
where $n$ is an integer, and the integral is along contour B. This is equation (\ref{eq:WKB}) in the main text. In the main text we transformed back to our coordinates.

\end{document}